\documentclass[a4paper,12pt]{article}
\pdfoutput=1

\usepackage{jheppub}
\usepackage[utf8]{inputenc}
\usepackage[english]{babel}

\usepackage{bm}
\usepackage{amsfonts}
\usepackage{mathbbol}
\usepackage{amsthm}
\usepackage{amssymb}
\usepackage{mathrsfs}
\usepackage{latexsym}
\usepackage{commath} 
\usepackage{float}
\usepackage{subfig}
\usepackage{subcaption}
\usepackage{comment}
\usepackage[makeroom]{cancel}

\begin{document}

\title
{Entanglement islands, fuzzballs and stretched horizons}

\author[a,b]{Dmitry S. Ageev} 
\author[a]{and Anastasia N. Zueva}

\affiliation[a]{Department of Mathematical Methods for Quantum Technologies, Steklov Mathematical
Institute of Russian Academy of Sciences, Gubkina str. 8, Moscow 119991, Russia}
\affiliation[b]{Institute for Theoretical and Mathematical Physics, Lomonosov Moscow State University, 119991 Moscow, Russia}

\emailAdd{ageev@mi-ras.ru}
\emailAdd{zueva@mi-ras.ru}

\abstract{
We study the implementation of the island prescription in fuzzball-inspired models of black holes. As a simplified setup, we model a fuzzball by replacing the event horizon with a reflecting boundary (stretched horizon). In the framework of two-dimensional model with such boundary, we analyze the dynamics of entanglement entropy. We find that the presence of the boundary modifies the behavior of the island saddle, and for a range of parameter values we observe the effect of blinking island found in \href{https://doi.org/10.1103/PhysRevD.111.026002}{arXiv:2311.16244}  which inevitably leads to the analogue of information paradox. 

We then extend the analysis to higher dimensions, incorporating both bulk and boundary contributions to the generalized entropy. The existence of island solutions is found to depend sensitively on the boundary conditions and the position of the stretched horizon, naturally leading to the absence of entanglement islands for a wide range of parameters.

Finally, we consider more ``realistic'' stringy fuzzball geometries, including superstrata and bubbling solutions, and estimate whether island solutions can arise in these backgrounds. The results indicate that the existence of islands depends on the behavior of the geometric area near the cap, and is not guaranteed in general.
}

\maketitle

\clearpage


\section{Introduction}

The black hole information problem \cite{Hawking:1975vcx,PhysRevD.14.2460,PhysRevLett.71.3743,Page:2013dx} has acquired, over the last few years, a striking new shape. On the one hand, the semiclassical analysis of evaporating black holes has undergone significant recent developments, with the advent of the island formula reproducing the Page curve and suggesting that the fine-grained entropy of Hawking radiation is controlled by quantum extremal surfaces in a gravitational region \cite{Penington:2019npb,Almheiri:2019hni,Almheiri:2019psf,Almheiri:2019qdq,Penington:2019kki}.On the other hand, string theory has long pointed toward a rather different perspective: the classical horizon may itself be a coarse-grained effective description, to be replaced in the microscopic theory by a horizonless cap or fuzzball state \cite{Mathur:2005zp,Bena:2007kg,Bena:2013dka}.  Each of these ideas addresses the same paradox.  Yet they do so from opposite sides.  Islands seem to teach us that semiclassical gravity knows more about unitarity than one might have expected, whereas fuzzballs suggest that the semiclassical interior may never have been fundamental to begin with.

This tension motivates the central question of this paper.  What becomes of the island prescription when the near-horizon region is replaced by a fuzzball-like cap? Equivalently, one may ask how much of the modern island framework is genuinely universal, and how much of it relies on geometric features that are special to black holes with horizons.  In the usual derivations, the quantum extremal surface sits in a geometry whose most salient feature is precisely the existence of a horizon and an associated redshift region.  In a fuzzball geometry, by contrast, there is no smooth interior to hide behind the horizon, because there is no horizon. It is not therefore evident a priori that the island's mechanism remains intact. The real issue is sharper: does a horizonless microstate geometry still admit a stable extremum of the generalized entropy, and if so, what geometric structure is responsible for it?

Our strategy is to approach this question in stages, moving from the simplest analytically tractable model to more realistic horizonless geometries. Islands in the context of two-dimensional gravity have been studied in \cite{Almheiri:2019yqk,Chen:2020uac,Chen:2020hmv,Almheiri:2020cfm}. And islands for a two-dimensional approximation of a Schwarzschild black hole were first used in \cite{Hashimoto:2020cas}, and further generalized and discussed in great detail in many scenarios, see for example \cite{Alishahiha:2020qza,Ling:2020laa,Matsuo:2020ypv,Karananas:2020fwx,Yu:2022xlh,PhysRevD.107.123509,Wang:2021woy,Kim:2021gzd,Lu:2021gmv,Yu:2021cgi,Ahn:2021chg,Cao:2021ujs,Azarnia:2021uch,Arefeva:2021kfx,He:2021mst,Yu:2021rfg,Arefeva:2022cam,Gan:2022jay,Azarnia:2022kmp,Anand:2022mla,Ageev:2022hqc,Ageev:2022qxv,Ageev:2023mzu}.
 The first step is deliberately crude.  We replace the horizon by a timelike reflecting wall placed just outside it. This ``stretched horizon'' is not meant to be a faithful representation of microstate geometry, but it does capture a common feature in many fuzzball-inspired models: the removal of the classical horizon and its replacement with boundary data in the near-horizon region. The advantage of this model is that it isolates the effect of removing the horizon while keeping the entropy calculation under analytic control.  In particular, the matter sector may be treated as a boundary conformal field theory on the two-dimensional conformal part of the geometry\cite{Rozali:2019day,Sully:2020pza,Geng:2021iyq,Ageev:2021ipd,Hollowood:2021wkw,Geng:2021mic,Suzuki:2022xwv,Hu:2022ymx,Hu:2022zgy,BasakKumar:2022stg,Afrasiar:2022ebi,Geng:2022dua}, and the effect of the wall can be followed explicitly through the entropy and the anomaly-induced stress tensor. 

This simple model already turns out to be surprisingly rigid. The wall does not simply modify the standard island picture; it can qualitatively alter it.  In the two-dimensional Schwarzschild problem, the no-island entropy still grows and then saturates, but the saturation value depends sensitively on the wall position and may exceed the thermodynamic entropy when the wall is placed sufficiently close to the horizon.  More strikingly, the island need not exist for all times.  For a range of parameters, the quantum extremal surface exists only during part of the evolution: the island appears, disappears, and may reappear again.  We call this effect ``blinking island'' by analogy with the similar effect found in the model of black hole cavity \cite{Ageev:2023hxe}.  During the interval in which the island is absent, the entropy of the radiation again exceeds the thermodynamic bound, and the information paradox reemerges. This indicates that once a boundary is introduced near the horizon, the island prescription is no longer guaranteed to provide a robust resolution of the paradox at all times.

The same conclusion can be seen from a different perspective when one studies the stress tensor.  A stretched horizon is not just an abstract boundary condition inserted into the entropy functional; it is an object that must be held in place.  Using the Herzog--Huang form \cite{Herzog:2017xha} of the boundary Weyl anomaly, we compute the renormalized stress tensor and the corresponding displacement operator for the matter sector in the presence of the wall.  This makes explicit the force density that the quantum fields exert on the boundary.  In the Schwarzschild case, the required external support diverges as the wall approaches the horizon.  Thus the closer one tries to move the wall toward the classical horizon, the harder it becomes to regard it as a natural or self-sustaining feature of the geometry.

We then turn to higher dimensions, where the question becomes less about exact solvability and more about mechanism.  Here one can no longer compute the full matter entropy exactly, but the near-horizon problem still has a clear approximation. Following the logic of recent work on large-angular-momentum islands and near-horizon mode extraction \cite{Hashimoto:2020cas,Matsuo:2020ypv,bousso2024islands}, we estimate the generalized entropy by combining the area term with the finite nonlocal part of the bulk entropy in a thin-slab approximation, and we supplement this with boundary-sensitive contributions motivated by entanglement in the presence of a wall. 

What emerges from this analysis is a simple geometric criterion.  A stable island requires the area term in the generalized entropy to grow sufficiently rapidly as one moves away from the cap.  In the near-cap region, this translates into a condition on the leading dependence of the area on proper distance.  If the area grows only linearly, the putative extremum of the generalized entropy is generically a local maximum rather than a minimum.  If the area grows quadratically, by contrast, the area term can balance the attractive nonlocal bulk entropy and support a stable quantum extremal surface.  In this sense, the existence of an island is not a universal consequence of horizonlessness, it depends on the detailed shape of the geometry near the true horizon.

This criterion provides a useful way to systematically classify fuzzball examples. For D1--D5 supertube and superstratum-like geometries, the cap and throat regions do not behave like the near-horizon region of Schwarzschild.  In the cap, the relevant area grows only linearly with proper distance, while in the long throat the area varies too slowly to produce a controlled local minimum of the generalized entropy. The corresponding stationary points are either absent or tend to the edge of the approximation area. By contrast, in five-dimensional bubbling geometries the warp factors combine to cancel the leading shrinking of the transverse sphere, so that near the cap the area is approximately constant plus a quadratic correction.  This is precisely the structure needed to support a stable island. 

Seen from this perspective, the main conceptual point is that islands and fuzzballs are not simply two complementary descriptions that automatically fit together.  The success of the island prescription in ordinary black hole backgrounds relies on specific geometric properties---a redshift region, a suitable competition between area growth and bulk entropy, and a clear identification of the radiation subsystem---that are not automatically reproduced in horizonless microstate geometries.  Some fuzzball models may admit a stable quantum extremal surface; others do not.  In models with a stretched horizon or a boundary, the island can even be transient.  The broader implication is that the mechanism of information recovery in a fuzzball description need not coincide with the semiclassical island mechanism, even if both are ultimately aimed at resolving the same paradox.

The rest of the paper develops this point systematically.  In Section \ref{sec:twhoDimens}, we analyze the two-dimensional Schwarzschild model with a reflecting wall and show explicitly how the island can blink in time.  In Section \ref{sec:stressTensor}, we compute the anomaly-induced stress tensor and the force required to hold the wall fixed near the horizon.  In Section \ref{sec:highDimens}, we generalize the static analysis to higher dimensions and formulate the corresponding extremization problem in a near-horizon approximation, including the role of boundary conditions.  In Section \ref{sec:qesFuzz}, we apply this logic to explicit fuzzball geometries, including D1--D5/superstratum and bubbling solutions, and identify the geometric conditions under which a stable island can or cannot exist.  We conclude in Section \ref{sec:conclusions} by discussing the implications of these examples for the relation between islands, stretched horizons, and fuzzball microphysics.

\section{Two-dimensional model and blinking island}\label{sec:twhoDimens}

\subsection{Entanglement entropy for a geometry with a boundary}

We begin by recalling the standard field-theoretic description of entanglement entropy.  Consider a quantum system in a pure state, described by a density matrix $\rho$.  Given a spatial region $R$, the degrees of freedom outside $R$ are traced out, producing the reduced density matrix $\rho_R$.  The entanglement entropy of $R$ is then the von Neumann entropy
\begin{equation}
    S(R)=-\text{Tr}\rho_R \log \rho_R .
\end{equation}
A convenient way to compute this quantity in quantum field theory is the replica trick.  In this construction, reviewed for example in \cite{Calabrese:2004eu}, one first computes the replicated partition function on an $n$-sheeted Euclidean geometry branched over the region $R$, and then analytically continues the answer to $n\rightarrow 1$.  Equivalently,
\begin{equation}
    S(R)=-\underset{n\rightarrow 1}{\lim}\partial_n (\text{Tr} \rho_R^n)
    =-\underset{n\rightarrow 1}{\lim}\partial_n\frac{Z_n(R)}{Z^n}.
\end{equation}
In two-dimensional conformal field theory this construction has a particularly useful local description.  The branch points of the replicated geometry may be represented by twist fields, whose correlation functions encode the replica partition function.  These twist fields implement the cyclic permutation symmetry among the replicated copies \cite{Calabrese:2009qy}.

For a two-dimensional Euclidean conformal field theory with a boundary, namely a BCFT, let the radiation region be a union of disjoint intervals
\begin{equation}
    R=[z_{a_1},z_{b_1}]\cup ... \cup [z_{a_m},z_{b_m}] .
\end{equation}
The entanglement entropy is then obtained from the corresponding twist-field correlator on the upper half-plane,
\begin{equation}
    S(R)=-\lim_{n\rightarrow 1}\partial_n
    \langle
    \phi(z_{a_1},\bar{z}_{a_1})
    \phi(z_{b_1},\bar{z}_{b_1})
    ...
    \phi(z_{a_m},\bar{z}_{a_m})
    \phi(z_{b_m},\bar{z}_{b_m})
    \rangle_\text{UHP}.
\end{equation}
Here $\phi$ denotes the twist operator.  It is a primary operator with conformal dimensions
\begin{equation}
    h_n=\bar{h}_n=\frac{c}{24}\left(n-\frac{1}{n}\right),
\end{equation}
where $c$ is the central charge of the matter theory.

In what follows we will need not only the correlators on the upper half-plane, but also their transformation under conformal maps.  Suppose that a geometry $\Omega$ is mapped to the upper half-plane by
\begin{equation}
    z:\Omega \rightarrow \text{UHP},\quad z=z(w), \quad \bar{z}=\bar{z}(\bar{w}).
\end{equation}
Since the twist operators are primary, their correlation functions transform as
\begin{equation}
    \begin{aligned}
        \langle \phi(w_1,\bar{w}_1)...\phi(w_m,\bar{w}_m) \rangle_\Omega
        =&\prod_{j=1}^m
        \left(\frac{d z}{d w} \right)^{h_n} \Bigg|_{w=w_j}
        \left( \frac{d\bar{z}}{d\bar{w}} \right)^{\bar{h}_n} \Bigg|_{\bar{w}=\bar{w}_j} \\
        &\times
        \langle \phi(z_1,\bar{z}_1)... \phi(z_m,\bar{z}_m) \rangle_\text{UHP}.
    \end{aligned}
\end{equation}
This relation will allow us to compute entropies in the black-hole geometry by first mapping the relevant two-dimensional Euclidean domain to the upper half-plane.

For the class of theories we will use below, namely $c$ copies of two-dimensional free massless Dirac fermions with perfectly reflecting boundary conditions \cite{CARDY1984514, Cardy:1986gw, Cardy:1989ir}, the entanglement entropy for an interval region $R$ takes the form \cite{Mintchev:2020uom, Rottoli_2023, Kruthoff:2021vgv}
\begin{equation}
    \begin{aligned}
        S(R)=&
        \frac{c}{3}\sum_{i,j=1}^n\ln |z_{a_i}-z_{b_j}|
        -\frac{c}{3}\sum_{i<j}^n \ln |z_{a_i}-z_{a_j}| |z_{b_i}-z_{b_j}|
        -c\,n \ln \varepsilon \\
        &+
        \frac{c}{6}\sum_{i,j=1}^n
        \ln |z_{a_i}-\bar{z}_{a_j}| |z_{b_i}-\bar{z}_{b_j}|
        - \frac{c}{6} \sum_{i,j=1}^n
        \ln |z_{a_i}-\bar{z}_{b_j}| |z_{b_i}-\bar{z}_{a_j}| ,
    \end{aligned}
\end{equation}
where $\varepsilon$ is the UV cutoff.  The first line is the bulk contribution, while the second line is the characteristic image-charge contribution induced by the boundary.

Our ultimate goal is to use this two-dimensional BCFT result as an effective description of matter in a higher-dimensional black-hole background.  Computing the exact matter entropy in a higher-dimensional curved spacetime is, in general, prohibitively difficult.  Following the strategy of \cite{Hashimoto:2020cas}, we exploit spherical symmetry and focus on geometries of the form
\begin{equation}
    ds^2 = e^{2\rho(w,\bar{w})}dw d\bar{w} + r^2d\Omega^2_d .
\end{equation}
The matter entropy is then approximated by that of the BCFT living on the two-dimensional conformal part of the metric,
\begin{equation}\label{eq:twoDimensionalPart}
    ds^2 = e^{2\rho(w,\bar{w})}dw d\bar{w}.
\end{equation}
and we assume that this reduced description captures the essential dynamics relevant for the entanglement calculation.

The remaining effect of curvature in two dimensions is encoded in a Weyl factor.  The entanglement entropy in flat space,
\begin{equation}
    ds^2=d w d\bar{w},
\end{equation}
and that in the curved metric \eqref{eq:twoDimensionalPart}
are related by the Weyl transformation
\begin{equation}\label{eq:EntCurved}
    S_m\bigg|_{ds^2=e^{2\rho(w,\bar{w})}dw d\bar{w}}
    =
    S_m\bigg|_{ds^2=d w d\bar{w}}
    +\frac{c}{6}\sum_{i=1}^m \ln e^{\rho(w_i,\bar{w}_i)}.
\end{equation}
Thus, in order to use the BCFT formulas above, we must first translate the two-sided black-hole geometry into a Euclidean two-dimensional geometry and then map it conformally to the upper half-plane.  We now turn to this geometric construction.

\subsection{Geometry}

We start from the Schwarzschild metric written in Kruskal coordinates,
\begin{equation}\label{eq:metricInKruskal}
    ds^2=-e^{2\rho(r)}dUdV+r^2d\Omega^2,\quad
    e^{2\rho(r)}=f(r)e^{-2\kappa r_*}.
\end{equation}
The Kruskal coordinates are defined by
\begin{equation}
        U=-\frac{1}{\kappa}e^{-\kappa(t-r_*)},
        \quad
        V=\frac{1}{\kappa}e^{\kappa(t+r_*)},
        \quad
        \kappa=\frac{1}{2 r_h}.
\end{equation}
For $r>r_h$, the tortoise coordinate is
\begin{equation}
    r_*(r)=r+ r_h \log \frac{r-r_h}{r_h}.
\end{equation}
It is also useful to introduce the timelike and spacelike Kruskal combinations
\begin{equation}
    T=\frac{1}{2}(U+V),\qquad X=\frac{1}{2}(V-U).
\end{equation}

  The Euclidean time is defined by $\tau=i t$ and has period $2\pi/\kappa$ and  equivalently, the Wick rotation of the Kruskal time is given by
\begin{equation}
    T=- i \mathcal{T}.
\end{equation}
Under this continuation the coordinate relations become
\begin{equation}
        \mathcal{X}=\pm e^{\kappa r_*}\frac{\cos \kappa \tau}{\kappa},
        \quad
        \mathcal{T}=\pm e^{\kappa r_*}\frac{\sin \kappa \tau}{\kappa}.
\end{equation}
The two signs distinguish the two exterior wedges and  in the Euclidean picture, the left half-plane, $\mathcal{X}<0$, and the right half-plane, $\mathcal{X}>0$, descend from the left and right wedges of the Lorentzian black hole.

The geometry becomes particularly transparent when we write
\begin{equation}
    \begin{aligned}
        \mathcal{X}^2+\mathcal{T}^2=\frac{e^{2\kappa r_*(r)}}{\kappa^2},
        \quad
        \tan(\kappa\tau)=\frac{\mathcal{T}}{\mathcal{X}}.
    \end{aligned}
\end{equation}
Surfaces of constant radial coordinate $r_0>r_h$ are circles of radius
\begin{equation}
    L=\frac{e^{\kappa r_*(r_0)}}{\kappa}
\end{equation}
centered at the origin $\mathcal{T}=\mathcal{X}=0$ and  it corresponds to the Lorentzian event horizon $r=r_h$.  By contrast, surfaces of constant Euclidean time are straight rays emanating from the same origin.

We now introduce a spherically symmetric reflecting boundary at radius $r_0>r_h$ in both exterior wedges of the analytically extended Schwarzschild geometry.  This boundary acts as a stretched horizon: it excises the true horizon and replaces it with a timelike wall.  In this sense it gives a simple effective model of a fuzzball-like cap.  After Wick rotation, the same configuration is represented by the exterior of a disk of radius
\begin{equation}
    L=\frac{e^{\kappa r^*_0}}{\kappa},
    \qquad r^*_0=r_*(r_0),
\end{equation}
with the disk boundary corresponding to the wall.

\begin{figure}[t!]
    \centering
    \includegraphics[width=1\linewidth]{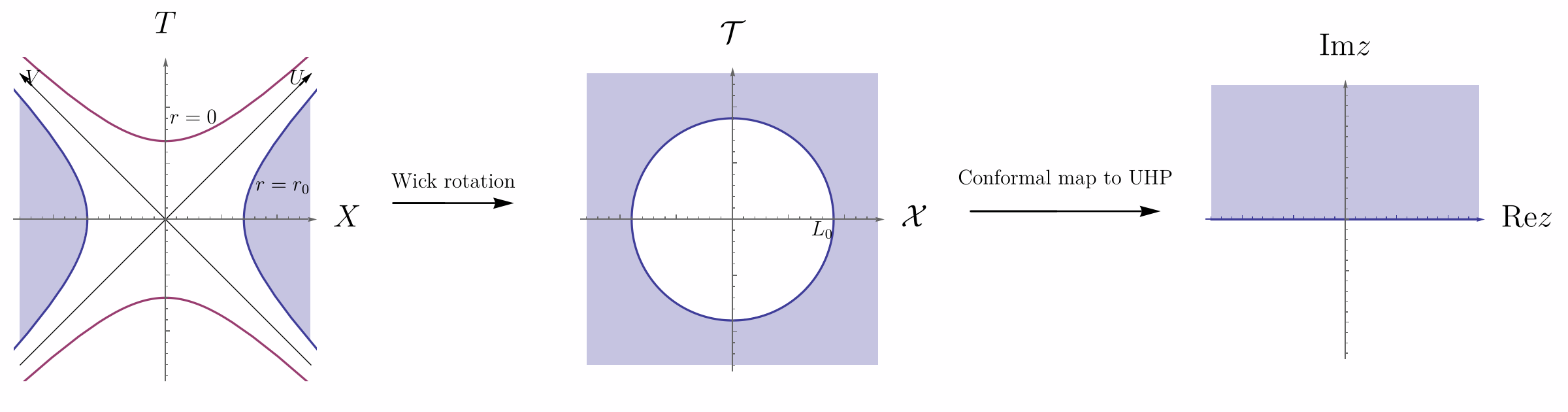}
    \caption{The transformation of a two-dimensional Schwarzschild black hole to Euclidean signature, followed by the conformal map to the upper half-plane.}
    \label{fig:MapPicture}
\end{figure}

Introducing complex coordinates
\begin{equation}
    w=\mathcal{X}+i \mathcal{T},\quad
    \bar{w}=\mathcal{X}-i \mathcal{T}.
\end{equation}
we perform the conformal  mapping of the exterior of the disk to the upper half-plane via
\begin{equation}
        z=-i\frac{L_0+w}{L_0-w},\quad
        \bar{z}=i\frac{L_0+\bar{w}}{L_0-\bar{w}}.
\end{equation}
After this map, the BCFT formulas of the previous subsection can be applied directly, with the Weyl factor in \eqref{eq:EntCurved} accounting for the curved background.  The sequence of transformations is summarized in Fig. \ref{fig:MapPicture}.

\subsection{Entanglement entropy without island}

We first compute the entropy in the absence of an island.  The radiation region $R$ is taken to consist of two semi-infinite intervals, one in each exterior wedge, beginning at the points
\begin{equation}\label{eq:pointsB}
    b_+=(r_b,t_b), \quad b_-=(r_b, -t_b),\quad r_b>r_0 .
\end{equation}
The Weyl factor appearing in \eqref{eq:EntCurved} and \eqref{eq:metricInKruskal} is the same at the two endpoints and is given by
\begin{equation}
    e^{\rho(b_+)}=e^{\rho(b_-)}
    =
    \sqrt{f(r_b)}e^{-\kappa r_*(r_b)} .
\end{equation}
Substituting these points into the BCFT expression and including the Weyl contribution, the entropy without an island becomes
\begin{equation}\label{eq:EntNoIsl}
    S_\text{noIsl}
    =
    \frac{c}{6}\ln\frac{4 f(r_b)\cosh^2\kappa t_b}{\kappa^2 \epsilon^2}
    +
    \frac{c}{6}\ln\frac{2\sinh^2 \kappa(r^*_b-r^*_0)}
    {\cosh 2\kappa(r^*_b-r^*_0)+\cosh 2\kappa t_b} .
\end{equation}
This expression displays two competing effects.  The first logarithm grows with the  time $t_b$, while the second term encodes the finite distance between the radiation endpoint and the reflecting wall.

At sufficiently late times,
\begin{equation}
    t_b \gg t_b^1,\quad t_b^1=r_b^*-r_0^*,
\end{equation}
the time dependence cancels between the two logarithms, and the entropy saturates to
\begin{equation}\label{eq:EntSaturate}
    S_\text{noIsl}
    =
    \frac{c}{6}\ln \frac{4 f(r_b)\sinh^2 \kappa(r^*_b-r^*_0)}
    {\kappa^2 \epsilon^2}.
\end{equation}
We compare this saturation value with the thermodynamic entropy of the two-sided black hole,
\begin{equation}\label{eq:EntBH}
    S^\text{therm}_\text{BH}=\frac{2\pi r_h^2}{G}.
\end{equation}
This is twice the Bekenstein--Hawking entropy \cite{Hawking:1975vcx, PhysRevD.7.2333} and  if the entanglement entropy of the radiation exceeds this thermodynamic entropy, the semiclassical description suggests a violation of unitary evolution for the combined system of the black hole and Hawking radiation.  We will refer to this case as the onset of the information paradox.

\begin{figure}[t!]
    \centering
    \includegraphics[width=0.7\linewidth]{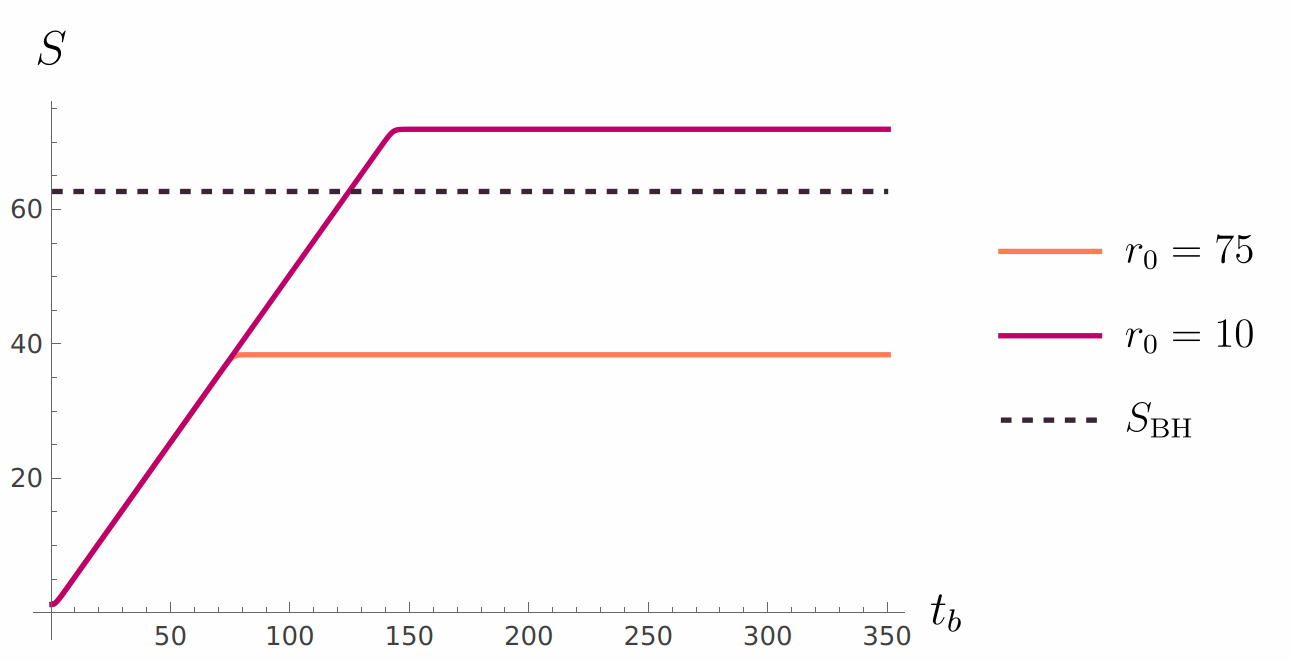}
    \caption{Comparison of the time dependence of the entanglement entropy for the simple fuzzball model without an island, computed for different wall positions $r_0$, with the thermodynamic entropy $S_\text{BH}$ shown by the \textit{dashed black} line.  The parameters are $r_b=150$, $c=3$, and $r_h=1$.  As the boundary $r_0$ approaches the horizon $r_h$, the entanglement entropy begins to exceed the thermodynamic entropy.}
    \label{fig:NoIslEnt}
\end{figure}

Fig. \ref{fig:NoIslEnt} illustrates this behavior.  As the wall is moved closer to the event horizon, the saturation value of the entropy increases and may exceed the thermodynamic entropy.  Thus, in the no-island case, the information paradox is most acute when the stretched horizon lies very close to the event horizon.

\subsection{Entanglement entropy with island}
The island prescription modifies the semiclassical computation by allowing part of the gravitational region to be included in the entanglement wedge of the radiation.  As advocated in \cite{Penington:2019npb} and \cite{Almheiri:2019psf}, and motivated by the quantum version of the Ryu--Takayanagi formula together with entanglement wedge reconstruction, the relevant entropy is the generalized entropy.  The prescription can be stated as the island formula \cite{Almheiri:2019hni}
\begin{equation}
    S(R)=\min \Biggl \{
    \underset{\partial I}{\text{ext}}
    \bigg[
    \frac{\text{Area}(\partial I)}{4 G}
    +S_\text{m}(R\cup I)
    \bigg]
    \Biggl\}.
\end{equation}
Here $\text{Area}(\partial I)$ is the area of the island boundary, while $S_\text{m}(R\cup I)$ is the ordinary matter entropy of the union of the radiation region and the island, computed in the fixed semiclassical geometry \cite{Penington:2019kki,Almheiri:2019qdq}.  The surface $\partial I$ is selected by extremizing the generalized entropy, and if several extrema exist, the one with minimal generalized entropy is chosen.

In the two-sided geometry we take the island $I$ to be a segment connecting two symmetric points,
\begin{equation}
    a_+=(r_a,t_a),\quad a_-=(r_a,-t_a),
\end{equation}
one in each exterior wedge and the generalized entropy functional is written as
\begin{equation}
    S_\text{gen}[I,R]=S_I(R)+S_I^b(R).
\end{equation}
The first contribution is the generalized entropy for the two-sided black hole with semi-infinite radiation regions, as obtained in \cite{Hashimoto:2020cas},
\begin{equation}
    \begin{aligned}
        S_I(R)=&
        \frac{2\pi r_a^2}{G}
        +\frac{c}{3}\ln
        \frac{4 \sqrt{f(r_a) f(r_b)}\cosh\kappa t_a \cosh \kappa t_b}
        {\kappa^2\epsilon^2}\\
        &+
        \frac{c}{3}\ln
        \frac{\cosh \kappa (r^*_a-r^*_b)-\cosh\kappa(t_a-t_b)}
        {\cosh \kappa (r^*_a-r^*_b)+\cosh\kappa(t_a+t_b)}.
    \end{aligned}
\end{equation}
The second contribution is the additional boundary-dependent term induced by the reflecting wall,
\begin{equation}\label{eq:EntBoundTerm}
    \begin{aligned}
        S_I^b(R)=&
        \frac{c}{6}\ln
        \frac{4\sinh^2\kappa(r^*_0-r^*_a)\sinh^2\kappa(r^*_0-r^*_b)}
        {\left(\cosh 2\kappa (r^*_0 - r^*_a) +
    \cosh 2\kappa t_a\right)
    \left(\cosh 2\kappa(r^*_0 - r^*_b) +
    \cosh 2\kappa t_b\right)}
    \\
    &+
    \frac{c}{3}\ln
    \frac{\cosh \kappa (2 r^*_0-r^*_a-r^*_b)+\cosh\kappa(t_a+t_b)}
    {\cosh\kappa (2 r^*_0 - r^*_a - r^*_b) - \cosh \kappa (t_a - t_b)}.
    \end{aligned}
\end{equation}
This boundary term vanishes in the limit $r_0\rightarrow r_h$, corresponding to the wall being pushed arbitrarily close to the event horizon.

The location of the island is determined by extremizing the generalized entropy with respect to both the radial coordinates and time of the island endpoints.  Thus we must solve
\begin{equation}
    \begin{cases}\label{eq:SystemDiff}
        \partial_{r_a}S_\text{gen}[I,R](r_a,t_a,r_b,t_b)=0
        \quad\text{or}\quad
        \partial_{r^*_a}S_\text{gen}[I,R](r_a,t_a,r_b,t_b)=0,\\
        \partial_{t_a}S_\text{gen}[I,R](r_a,t_a,r_b,t_b)=0.
    \end{cases}
\end{equation}
For convenience, we use $r^*_a=r_*(r_a)$ as the radial variable.

The exact derivatives of the generalized entropy functional are
\begin{equation}
    \begin{aligned}
        \frac{\partial S_\text{gen}}{\partial t_a}
        &=
        \frac{c}{3}\kappa
        \Bigg[
        \frac{\sinh (\kappa (t_a-t_b))}
        {\cosh(\kappa (2 r^*_0-r^*_a-r^*_b))-\cosh (\kappa(t_a-t_b))}
        \\
        &+
        \frac{\sinh (\kappa (t_a-t_b))}
        {\cosh (\kappa(t_a-t_b))-\cosh (\kappa(r^*_a-r^*_b))}
        \\
        &+
        \frac{\sinh (\kappa(t_a+t_b))}
        {\cosh (\kappa (2 r^*_0-r^*_a-r^*_b))+\cosh (\kappa(t_a+t_b))}
        \\
        &-
        \frac{\sinh(2 \kappa t_a)}
        {\cosh (2 \kappa(r^*_0-r^*_a))+\cosh (2 \kappa t_a)}
        \\
        &-
        \frac{\sinh (\kappa(t_a+t_b))}
        {\cosh (\kappa(r^*_a-r^*_b))+\cosh (\kappa(t_a+t_b))}
        +\tanh (\kappa t_a)
        \Bigg]
    \end{aligned}
\end{equation}
and
\begin{equation}
    \begin{aligned}
        \frac{\partial S_\text{gen}}{\partial r^*_a}
        &=
        \frac{2 \pi W\left(e^{2 \kappa r^*_a-1}\right)}{G \kappa}
        +\frac{c\kappa}{3 \left(W\left(e^{2 \kappa r^*_a-1}\right)+1\right)^2}
        \\
        &+
        \frac{c}{3} \kappa
        \Bigg[
        \frac{\sinh (\kappa (2r^*_0-r^*_a-r^*_b))}
        {\cosh(\kappa (2 r^*_0-r^*_a-r^*_b))-\cosh(\kappa (t_a-t_b))}
        \\
        &-
        \frac{\sinh(\kappa (2 r^*_0-r^*_a-r^*_b))}
        {\cosh(\kappa (2 r^*_0-r^*_a-r^*_b))+\cosh(\kappa (t_a+t_b))}
        \\
        &+
        \frac{\sinh (2\kappa (r^*_0-r^*_a))}
        {\cosh (2\kappa (r^*_0-r^*_a))+\cosh (2 \kappa t_a)}
        -\coth (\kappa(r^*_0-r^*_a))
        \\
        &+
        \frac{\sinh (\kappa(r^*_a-r^*_b))}
        {\cosh (\kappa(r^*_a-r^*_b))-\cosh (\kappa(t_a-t_b))}
        \\
        &-
        \frac{\sinh (\kappa(r^*_a-r^*_b))}
        {\cosh (\kappa(r^*_a-r^*_b))+\cosh (\kappa(t_a+t_b))}
        \Bigg].
    \end{aligned}
\end{equation}
To understand the structure of these equations analytically, we study them in the near-horizon and late-time regime.  The assumptions are as follows:
\begin{itemize}
    \item The wall and the island are both close to the horizon, so that
    \begin{equation}
        r^*_0,r^*_a<0\quad\text{and}\quad \kappa|r^*_0-r^*_a|\gg 1.
    \end{equation}

    \item The radiation endpoint is sufficiently far from the boundary,
    \begin{equation}
        r^*_b\gg r^*_0,
    \end{equation}
    so that
    \begin{equation}
        |2 r^*_0-r^*_a -r^*_b|\gg |t_a-t_b|.
    \end{equation}

    \item We work at late times,
    \begin{equation}
        \kappa (t_a+t_b)\gg 1,
        \qquad
        2\kappa t_a\gg 1.
    \end{equation}

    \item In the radial derivative we use the leading behavior of the principal branch $W_0(x)$ of the Lambert function near $x=0$, since
    \begin{equation}
        e^{-1+2\kappa r^*_a}\ll 1
        \quad \text{with } r^*_a<0.
    \end{equation}
\end{itemize}
Under these assumptions the derivatives simplify to
\begin{equation}
    \begin{aligned}
        \frac{\partial S_\text{gen}}{\partial r^*_a}
        &\approx
        \frac{2 \pi  e^{2 \kappa r^*_a-1}}{G \kappa}
        -\frac{2 c}{3}\kappa e^{2 \kappa r^*_a-1}
        \\
        &+
        \frac{c}{3} \kappa
        \Bigg[
        1+
        \frac{e^{-\kappa (2 r^*_0-r^*_a-r^*_b)}}
        {e^{-\kappa(2 r^*_0-r^*_a-r^*_b)}+e^{\kappa(t_a+t_b)}}
        -
        \frac{e^{-2 \kappa(r^*_0-r^*_a)}}
        {e^{-2 \kappa(r^*_0-r^*_a)}+e^{2 \kappa t_a}}
        \\
        &+
        \frac{e^{-\kappa(r^*_a-r^*_b)}}
        {e^{-\kappa(r^*_a-r^*_b)}+e^{\kappa(t_a+t_b)}}
        -
        \frac{e^{-\kappa(r^*_a-r^*_b)}}
        {e^{-\kappa(r^*_a-r^*_b)}-2 \cosh(\kappa(t_a-t_b))}
        \Bigg]
    \end{aligned}
\end{equation}
and
\begin{equation}
    \begin{aligned}
        \frac{\partial S_\text{gen}}{\partial t_a}
        &\approx
        \frac{c}{3} \kappa
        \Bigg[
        1+
        \frac{e^{\kappa(t_a+t_b)}}
        {e^{\kappa (-2 r^*_0+r^*_a+r^*_b)}+e^{\kappa(t_a+t_b)}}
        -
        \frac{e^{2 \kappa t_a}}
        {e^{2 \kappa(r^*_a-r^*_0)}+e^{2 \kappa t_a}}
        \\
        &-
        \frac{e^{\kappa(t_a+t_b)}}
        {e^{-\kappa(r^*_a-r^*_b)}+e^{\kappa(t_a+t_b)}}
        +
        \frac{\sinh(\kappa(t_a-t_b))}
        {\cosh(\kappa(t_a-t_b))-\frac{1}{2} e^{-\kappa(r^*_a-r^*_b)}}
        \Bigg].
    \end{aligned}
\end{equation}

\begin{figure}[t!]
    \centering
    \subfloat[]
    {\includegraphics[width=0.49\linewidth]{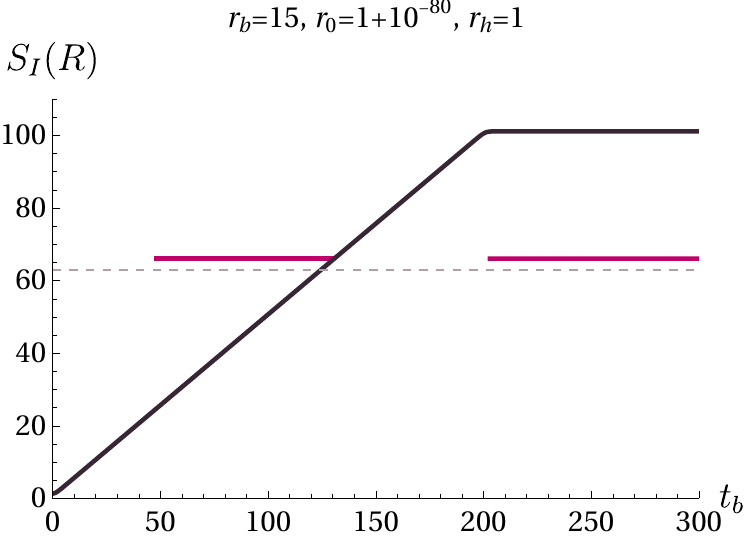}}
    \subfloat[]
    {\includegraphics[width=0.49\linewidth]{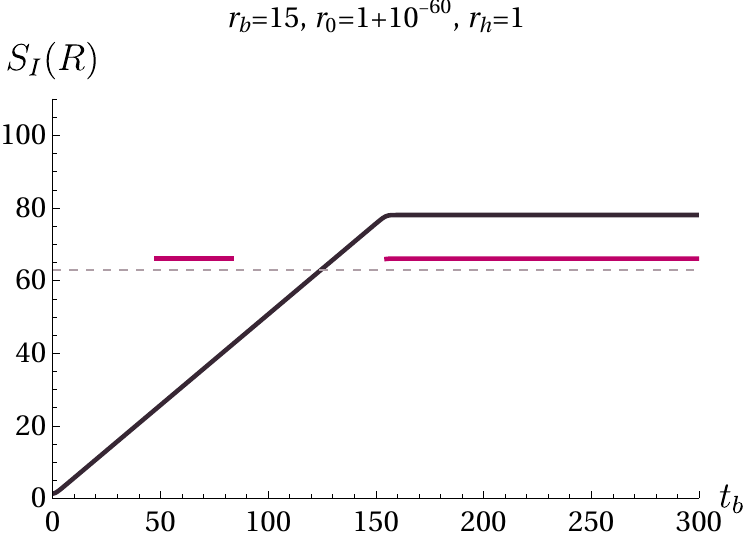}}

    \subfloat[]
    {\includegraphics[width=0.49\linewidth]{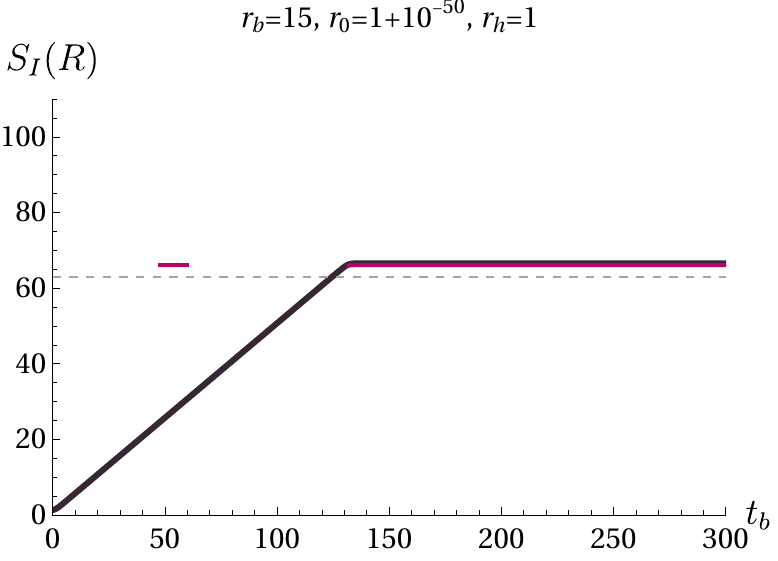}}
    \subfloat[]
    {\includegraphics[width=0.49\linewidth]{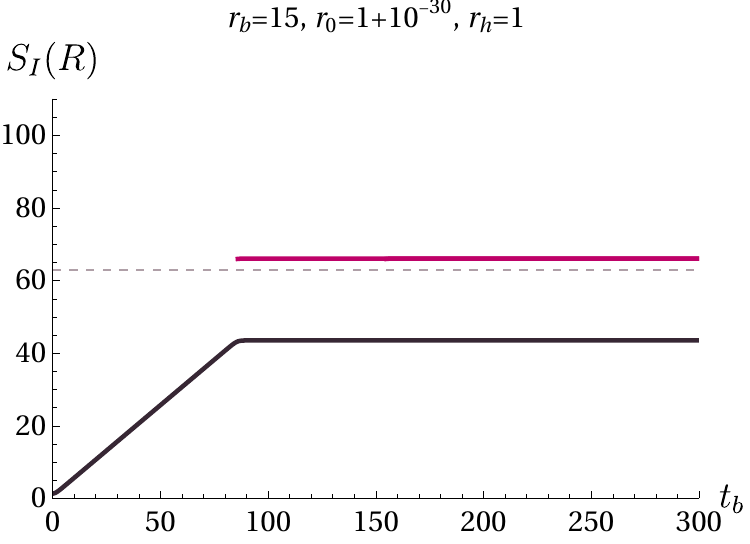}}
    \caption{Blinking island behavior for different positions of the boundary $r_0$.}
    \label{fig:BlinkingIslandPosition}
\end{figure}

The  behavior of the island is summarized in Fig. \ref{fig:BlinkingIslandPosition}.  As the wall is moved away from the horizon, the island begins to appear at earlier times.  For a sufficiently distant wall, the island ceases to be transient and becomes permanent.

\begin{figure}[t!]
    \centering
    \subfloat[$r_b=15$]
    {\includegraphics[width=0.4\linewidth]{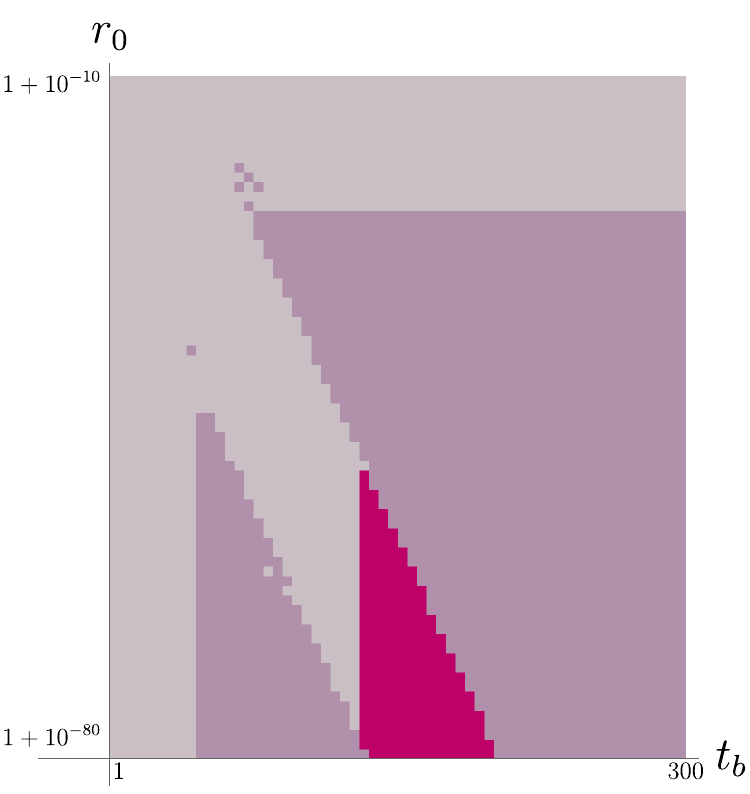}}
    \subfloat[$r_b=5$]
    {\includegraphics[width=0.4\linewidth]{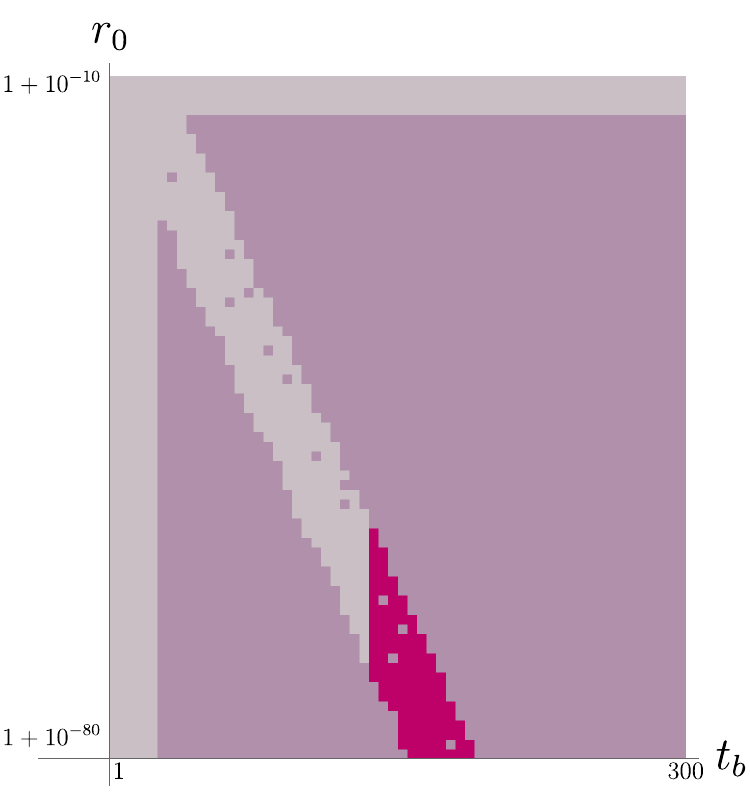}}
    \caption{Existence of an island and the appearance of the information paradox as functions of the wall position $r_0$ and the time $t_b$. \textit{Gray} denotes the absence of an island, \textit{purple} denotes the region where an island exists, and \textit{pink} denotes the region where no island exists and the information paradox arises.  Here $r_h=1$.}
    \label{fig:MatrixPlot}
\end{figure}

A broader picture is shown in Fig. \ref{fig:MatrixPlot}.  The plot separates the parameter space into regions where an island exists and regions where the no-island entropy exceeds the thermodynamic entropy.  In this way it displays, for each wall position and boundary time, whether the island prescription resolves the information paradox.
At this point, it is important to make a remark about the scale. As shown in Fig.~\ref{fig:MatrixPlot}, the island solution exists only when the wall is much closer to the horizon than the Planck length.  For the parameters used here, $r_h=1$ and $G=0.01$, the Planck proper distance gives
\begin{equation}
    l_p=\sqrt{G}=2\sqrt{r_h(r_\text{pl}-r_h)}
    \Rightarrow r_\text{pl}-r_h=0.0025\gg r_0-r_h .
\end{equation}
Thus the wall must be placed parametrically inside the Planck-scale radial separation from the horizon in order for this particular island saddle to exist.

\subsection*{Stress tensor from the  boundary anomaly}\label{sec:stressTensor}

We now turn to the stress tensor of the two-dimensional conformal field theory in the presence of the reflecting wall.  The purpose of this section is to make explicit how the boundary contribution to the Weyl anomaly is represented in the expectation value of the stress tensor.

We consider a two-dimensional CFT with central charge $c$ on a static background with metric
\begin{align}\label{eq:metricApp}
ds^2 = - f(r)\,dt^2 + \frac{dr^2}{f(r)} \, , \qquad r \ge r_0 \, .
\end{align}
The wall is located at fixed radius $r=r_0$, and the matter theory is defined only in the exterior region.  Throughout this discussion the geometry is treated as nondynamical, and all stress tensors are understood as renormalized expectation values.

The renormalized stress tensor is defined in the usual way by varying the generating functional $W[g;\mathcal B]$ with respect to the metric. Possible boundary terms, associated with the boundary condition $\mathcal{B}$, will not play an explicit role in what follows.

In two dimensions, renormalization produces the conformal or Weyl anomaly: the trace of the stress tensor does not vanish even in a conformal theory.  On a manifold $M$ with boundary $\partial M$, the anomaly takes the compact Herzog--Huang form \cite{Herzog:2017xha}
\begin{align}\label{eq:HerzogHuang}
\langle T^\mu{}_\mu \rangle
=
\frac{c}{24\pi}\Big(R + 2K\,\delta(x_\perp)\Big) \, .
\end{align}
Here $R$ is the bulk Ricci scalar, $K$ is the trace of the extrinsic curvature of $\partial M$ computed with the outward-pointing unit normal, and $x_\perp$ is the proper distance transverse to the boundary, so that $\delta(x_\perp)$ is supported at $\partial M$.  We work within this form of the anomaly.  In particular, we ignore scheme-dependent total-derivative terms and do not include an additional boundary cosmological constant.

The details of the derivation are given in Appendix \ref{app:stressTensor}.  For the metric \eqref{eq:metricApp}, the trace anomaly becomes
\begin{align}\label{eq:anomalySection}
\langle T^\mu{}_\mu \rangle
&=
\frac{c}{24\pi}\Big(-f''(r) - f_0'\,\delta(r-r_0)\Big) \, .
\end{align}
The first term is the usual bulk curvature contribution.  The second term is localized at the wall and is the direct image of the boundary term $K\,\delta(x_\perp)$ in \eqref{eq:HerzogHuang}.

A minimal way to construct this boundary-localized anomaly is to add a stress tensor localized at the wall and tangent to the boundary. We choose
\begin{align}\label{eq:stressBound}
\langle T_{tt}\rangle_{\rm bdry}
=
\frac{c}{24\pi} f_0 f_0'\,\delta(r-r_0)\, ,
\qquad
\langle T_{rr}\rangle_{\rm bdry}
=
\langle T_{tr}\rangle_{\rm bdry}
=
0 \, .
\end{align}
Its contribution to the trace is $g^{tt}\langle T_{tt}\rangle_{\rm bdry}$, which precisely reproduces the boundary part of \eqref{eq:anomalySection}.  The complete stress tensor is therefore obtained by combining the smooth bulk solution for $r>r_0$ with the $\delta$-function term \eqref{eq:stressBound} at the wall.

The bulk stress tensor is fixed, up to constants, by the Ward identities
\begin{equation}
    \nabla_\mu\langle T^\mu{}_\nu\rangle=0 .
\end{equation}
Solving these identities together with the bulk trace anomaly gives
\begin{align}
\langle T_{tt}\rangle_{\rm bulk}
&=
\Omega + \frac{c}{24\pi}f f''
- \frac{c}{96\pi}\left(f'\right)^2 \, ,
\nonumber\\
\langle T_{rr}\rangle_{\rm bulk}
&=
\frac{1}{f^2}
\left(
\Omega - \frac{c}{96\pi}\left(f'\right)^2
\right) \, ,
\nonumber\\
\langle T_{tr}\rangle_{\rm bulk}
&=
\frac{\Phi}{f} \, .
\end{align}
The constant $\Phi$ measures the energy flux.  A reflecting wall imposes vanishing flux through the boundary, namely $T^r{}_t(r_0)=0$, and therefore
\begin{equation}
    \Phi=0.
\end{equation}
Because the wall removes the horizon from the matter region, there is no regularity condition at $r=r_h$ that would otherwise fix the remaining integration constant $\Omega$.  Thus, formally, one free parameter remains.  Since the wall is assumed to sit very close to the true horizon, we fix this parameter by requiring the state to be thermal at infinity.

\subsection*{Fixing $\Omega$: KMS state and Tolman temperature}

To determine $\Omega$, we specify the asymptotic state.  We impose a KMS condition with respect to the Killing time,
\begin{equation}
    t\sim t+ i\beta_\infty,
\end{equation}
where $\beta_\infty$ is the inverse temperature measured at infinity.  Since $f(r)\rightarrow 1$ as $r\rightarrow\infty$, the metric becomes asymptotically flat.  In this region we introduce null coordinates
\begin{gather}
    u=t-x,\quad v=t+x,\\ \nonumber
    u\sim u+ i\beta_\infty,\quad v\sim v+i\beta_\infty.
\end{gather}
The asymptotic stress tensor is therefore that of a two-dimensional CFT in a thermal state.

The standard conformal transformation law gives \cite{Polchinski:1998rq}
\begin{equation}
    \langle T_{uu}\rangle =-\frac{c}{24 \pi}\{U,u\},
\end{equation}
where $\{U,u\}$ is the Schwarzian derivative.  A thermal state is obtained through the exponential map
\begin{equation}
    U(u) = -\frac{1}{a}e^{-a u},\quad
    V(v)=\frac{1}{a} e^{a v}, \quad
    a=\frac{2\pi}{\beta_\infty} = 2\pi T_\infty \, .
\end{equation}
It follows that
\begin{equation}
    \langle T_{uu}\rangle
    =
    \langle T_{vv}\rangle
    =
    \frac{c}{24 \pi}\frac{a^2}{2},
    \quad
    \langle T_{tt}\rangle_{\infty}
    =
    \langle T_{uu}\rangle + \langle T_{vv}\rangle
    =
    \frac{c\; \pi}{6} T_\infty^2.
\end{equation}
By definition, the integration constant $\Omega$ is the asymptotic energy density in the static frame.  Hence
\begin{align}
\Omega
=
\lim_{r\to\infty}\langle T_{tt}\rangle_{\rm bulk}
=
\frac{\pi c}{6}T_\infty^2 \, .
\end{align}
It is often more useful to parametrize the state by the proper temperature measured at the wall. A static observer at radius $r$ has proper time $d\tau=\sqrt{f(r)}\,dt$, so the local temperature obeys the Tolman relation \cite{PhysRev.35.904,PhysRev.36.1791}
\begin{align}
T_{\rm loc}(r)\sqrt{f(r)} = T_\infty \, .
\end{align}
Defining the proper wall temperature $T_0 \equiv T_{\rm loc}(r_0)$ gives $T_\infty=\sqrt{f_0}\,T_0$, and therefore
\begin{align}\label{eq:OmegaAppendix}
\Omega
=
\frac{\pi c}{6} f_0 T_0^2 \, ,
\qquad
T_{\rm loc}(r)
=
T_0\sqrt{\frac{f_0}{f(r)}} \, .
\end{align}

Finally, using the boundary Ward identity
\begin{equation}
    \nabla_\mu T^{\mu x}=\delta(x)D
\end{equation}
as described in Appendix \ref{app:displacement}, one obtains the displacement expectation value for the Schwarzschild metric,
\begin{align}
D
=
\frac{\pi c}{6}T_0^2
+
\frac{c}{24\pi}a_0^2
=
\frac{\pi c}{6}T_0^2
+
\frac{c}{96\pi}
\frac{r_h^2}{r_0^4\big(1-\frac{r_h}{r_0}\big)}\, .
\end{align}
This quantity is the normal force density exerted by the matter sector on the wall.  To keep the wall fixed at radius $r_0$, an external force must be applied in the opposite direction.  In the strict conformal framework of Herzog--Huang, with no additional boundary tension term, both contributions are non-negative for thermal states, and the wall must therefore be externally supported.

Two features of this result are worth emphasizing.  First, the thermal part of the displacement is governed by the Tolman temperature,
\begin{equation}
    T_{\rm loc}(r_0)=\frac{T_\infty}{\sqrt{1-r_h/r_0}}.
\end{equation}
Thus, at fixed $T_\infty$, it blue-shifts as the wall approaches the horizon.  Second, even at fixed proper wall temperature $T_0$, the geometric contribution to $D$ diverges as $r_0\to r_h$.  This divergence reflects the large proper acceleration, directed toward increasing $r$, required to hold the wall static near the horizon.



\section{Stretched horizon in higher dimensional black hole}\label{sec:highDimens}

We now extend the discussion to higher dimensions.  In two dimensions the matter entropy can be computed explicitly using conformal symmetry and BCFT methods. In higher dimensions, the entropy cannot usually be computed exactly, so we instead extract the leading terms that determine the island saddle.

We study the generalized entropy
\begin{equation}
    S_\text{gen}(R\cup I)
    =
    \frac{A(\partial I)}{4G}
    +
    S_\text{bulk}(R\cup I),
\end{equation}
in the near-horizon region of a $d$-dimensional Schwarzschild black hole. This regime is particularly useful because, near the horizon, the geometry approximately factorizes into two-dimensional Rindler space and a transverse $(d-2)$-sphere whose radius varies slowly with the proper distance from the horizon. This allows us to parametrize the island location by the proper distance $\rho_I$ from the horizon. The area term can then be expanded analytically, while the bulk entropy can be estimated by a thin-slab approximation for the region between the island boundary at $\rho_I$ and the radiation cutoff at $\rho_R$.

Our purpose is not to compute the full entropy exactly, but rather to identify the mechanism responsible for the appearance of the island saddle. The extremization problem is controlled by the interplay between the area term, which is sensitive to the variation of the transverse sphere with $\rho_I$, and the finite nonlocal contribution to the bulk entropy, which depends on the separation $\rho_R-\rho_I$. We then include the effect of a reflecting boundary at $\rho_0$ and analyze how the associated boundary contribution modifies the existence and stability of the island saddle.

The $d$-dimensional Schwarzschild black hole mentioned above has a metric
\begin{equation}
    ds^2 =- f(r) dt^2 +\frac{1}{f(r)}dr^2+ r^2 d\Omega^2_{d-2},\quad f(r)=1-\frac{r_h^{d-3}}{r^{d-3}},
\end{equation}
where the horizon is located at $r=r_h$.

Near the horizon, where $r-r_h \ll r_h$, we introduce the proper distance
\begin{equation}
    \rho=2\sqrt{\frac{r_h(r-r_h)}{d-3}}.
\end{equation}
and using it the metric takes the form
\begin{equation}
    ds^2=-\frac{(d-3)^2}{4 r_h^2}\rho^2 dt^2+d\rho^2 +\left(r_h+\frac{d-3}{4 r_h}\rho^2\right)^2 d\Omega^2_{d-2}.
\end{equation}
As before, the location of the island is determined by extremizing the generalized entropy
\begin{equation}
    \frac{d}{d\rho_I}S_\text{gen}(R\cup I)=0.
\end{equation}
We restrict to a static setup and the boundary $\partial I$ is a sphere of fixed radius on the $t=0$ slice. 
In the near-horizon regime $\rho_I \ll r_h$, the derivative of the area term is
\begin{equation}\label{eq:SAreaBousso}
    \frac{d A(\partial I)}{d\rho_I}=\frac{(d-2)(d-3)\pi^{(d-1)/2}}{\Gamma ((d-1)/2)}r_h^{d-4}\rho_I.
\end{equation}
To estimate the bulk entropy, we closely follow  \cite{Hashimoto:2020cas, Matsuo:2020ypv,bousso2024islands} in which the region $R$ is described as a set of points with $t=0$ and $\rho>\rho_R$. Using thin-slab approximation with transverse area \mbox{$A_h = r_h^{d-2}$}, the entropy of $R \cup I$ can be written as
 \begin{equation}\label{eq:SBulkBousso}
     S_\text{bulk}(R\cup I)=\frac{A_h}{\epsilon_I^{d-2}}+\frac{A_h}{\epsilon_R^{d-2}}-\kappa\frac{A_h}{(\rho_R-\rho_I)^{d-2}},
 \end{equation}
where $\epsilon_I$ and $\epsilon_R$ regulate short-distance divergences near the boundaries of $I$ and $R$. The coefficient $\kappa$ is a universal constant that depends on the specifics of the bulk fields and the spacetime dimension. In the case of a single free, massless scalar field in four-dimensional spacetime, $\kappa \approx 0.0049$ \cite{Srednicki:1993im,Lohmayer:2009sq}.

\begin{figure}[t!]
    \centering
    {\includegraphics[width=0.8\linewidth]{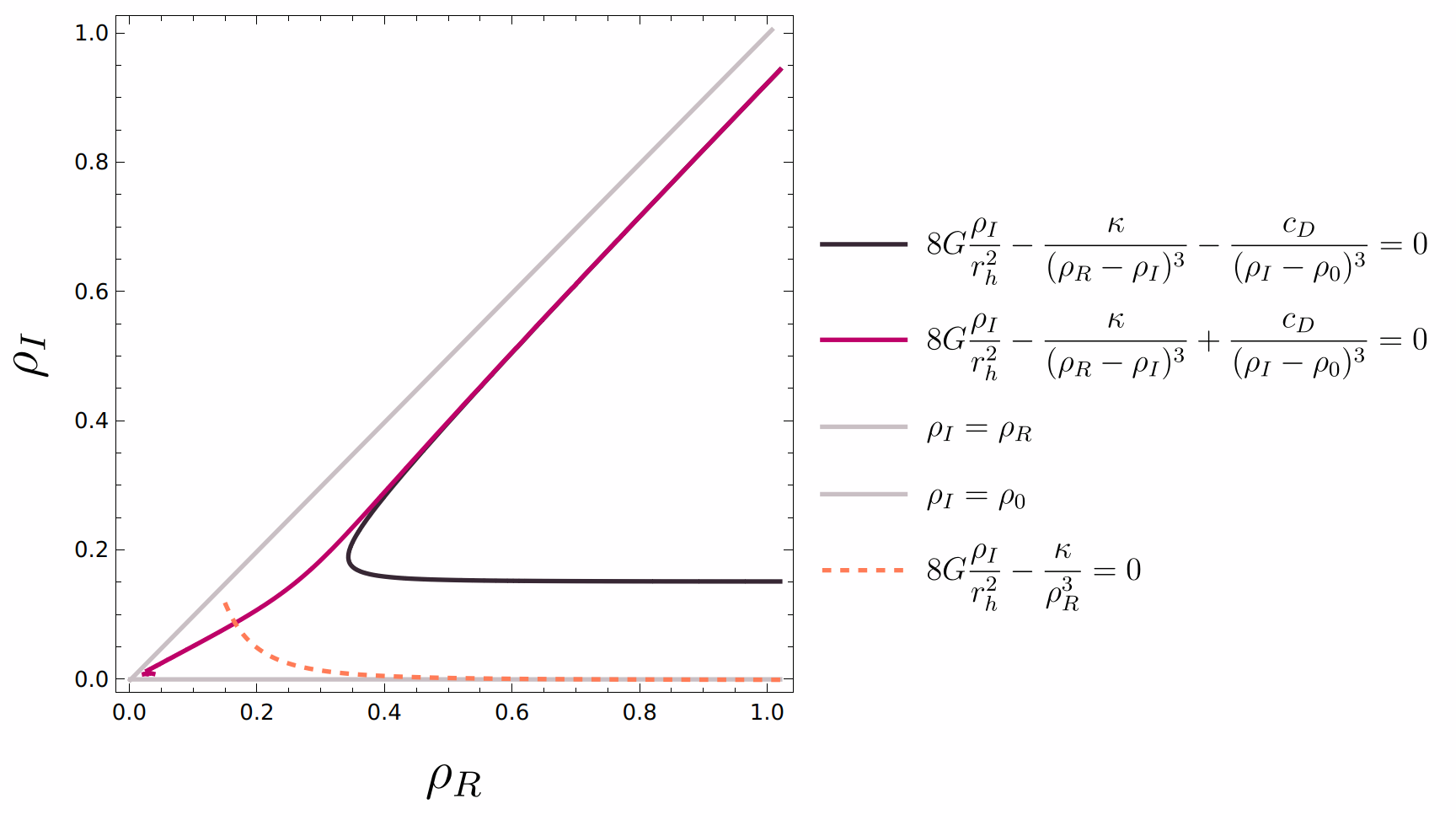}}
    \caption{The solution of \eqref{eq:fullEqOnIslandLocation} with Neumann \textit{(black line)} and Dirichlet \textit{(pink line)} boundary conditions, $d=4,\, r_h=1,\,\kappa = 4.9\cdot 10^{-3},c_d=6.63\cdot 10^{-3}, \rho_0=10^{-3}$.
    }\label{fig:FullDirAndNeu}
\end{figure}

In this work, we consider the effect of introducing a boundary at $\rho_0$ near the horizon. The contribution of a boundary to the entanglement entropy has been analyzed in \cite{berthiere2016boundary} for the case of a strip directly adjacent to the boundary, where the leading behavior scales as \mbox{$\sim 1/L^{d-2}$}. However, for a more general configuration in which the entangling region does not directly adjoin the boundary, i.e. a slab extending between $\rho_I$ and $\rho_R$ with \mbox{$\rho_0<\rho_I<\rho_R$}, a complete expression for the boundary contribution has not yet been determined and may involve additional non-local terms. In the absence of a general result, we adopt a simple ansatz motivated by the structure of the adjacent-strip case and assume that the boundary contribution can be approximated by a sum of terms associated with the distances to the boundary. Concretely, we take
\begin{equation}\label{eq:SBoundSol}
    S_\text{boundary}=\pm c_d A_h\left(\frac{1}{(\rho_I-\rho_0)^{d-2}}+\frac{1}{(\rho_R-\rho_0)^{d-2}}\right),
\end{equation}
where the plus sign corresponds to Neumann boundary conditions and the minus sign to Dirichlet boundary conditions. The prefactor for massless field is given by
\begin{equation}
    c_d =\frac{\Gamma (d/2)}{6(d-2)(4\pi)^{(d-2)/2}}.
\end{equation}
Only the first term depends on $\rho_I$  giving us 
\begin{equation}
    \frac{d S_\text{boundary}}{d\rho_I}=\mp (d-2)\frac{c_d A_h}{(\rho_I-\rho_0)^{d-1}},
\end{equation}
 and combining all contributions and dividing by $(d-2)A_h$, the extremization condition for the generalized entropy becomes
\begin{equation}\label{eq:fullEqOnIslandLocation}
    \frac{d S_\text{gen}}{d\rho_I}=\frac{(d-3)}{8G}\frac{\rho_I}{r_h^2}-\frac{\kappa}{(\rho_R-\rho_I)^{d-1}}\mp\frac{c_d }{(\rho_I-\rho_0)^{d-1}}=0.
\end{equation}
In Fig.~\ref{fig:FullDirAndNeu} we show the solutions of equation \eqref{eq:fullEqOnIslandLocation} with Dirichlet and Neumann boundary conditions, as well as an approximate solution \eqref{eq:Bousso}, which we will discuss later.

$\,$

To begin with, let's consider the case when $\kappa \gg c_D$, which essentially corresponds to the absence of the wall described in \cite{He:2021mst}. In this case the equation \eqref{eq:fullEqOnIslandLocation} simplifies to
\begin{equation}\label{eq:BoussoFull}
    \rho_I=\frac{8G\kappa\, r_h^2}{(d-3)(\rho_R- \rho_I)^{d-1}}.
\end{equation}
If we also assume that \mbox{$\rho_I\ll \rho_R$}, then we can get an approximation of the island location
\begin{equation}\label{eq:Bousso}
    \rho_I=\frac{8G\kappa\, r_h^2}{(d-3)\rho_R^{d-1}}, \quad \text{for } d=4\quad \rho_I = \frac{8 \kappa\, l_p^2\, r_h^2 }{\rho_R^{3}}.
\end{equation}
\subsection*{Dirichlet boundary conditions}

First let us consider the Dirichlet boundary conditions corresponding to
\begin{equation}
    \frac{d S_\text{gen}}{d \rho_I}=\frac{(d-3)}{8G}\frac{\rho_I}{r_h^2}-\frac{\kappa}{(\rho_R-\rho_I)^{d-1}}+\frac{c_d }{(\rho_I-\rho_0)^{d-1}}.
\end{equation}
We consider a scenario where the island is located very close to the boundary and the radiation region is far away, i.e. 
\begin{equation}
    \rho_I\approx\rho_0, \quad \delta\equiv \rho_I-\rho_0\ll \rho_R-\rho_I\approx\rho_R-\rho_0\equiv L.
\end{equation}
Then the equation for $\delta$ would be
\begin{equation}
    0=\frac{(d-3)}{8G}\frac{\rho_0}{r_h^2}-\frac{\kappa}{L^{d-1}}+\frac{c_d }{\delta^{d-1}}.
\end{equation}
with the solution in terms of $\rho_I$ given by
\begin{equation}
    \rho_I=\rho_0+\left(\frac{c_d}{\frac{\kappa}{(\rho_R-\rho_0)^{d-1}}-\frac{(d-3)}{8G}\frac{\rho_0}{r_h^2}} \right)^{\frac{1}{d-1}},
\end{equation}
where we assume \mbox{$\frac{\kappa}{(\rho_R-\rho_0)^{d-1}}>\frac{(d-3)}{8G}\frac{\rho_0}{r_h^2}$}.
If one imposes  that  \mbox{$\frac{\kappa}{(\rho_R-\rho_0)^{d-1}}\gg \frac{(d-3)}{8G}\frac{\rho_0}{r_h^2}$}, then the solution is 
\begin{equation}
    \rho_I\approx \rho_0 +\left(\frac{c_d}{\kappa}\right)^{\frac{1}{d-1}}(\rho_R-\rho_0), \quad \frac{c_d}{\kappa}\ll1.
\end{equation}
Note that the approximate solutions found using the Dirichlet boundary condition correspond to local maxima of the generalized entropy. At the ends of the interval $(\rho_0,\rho_R)$, which could be considered a trivial island solution, the entropy diverges. Therefore, we can say that Dirichlet boundary conditions prevent an island solution.

\subsection*{Neumann boundary conditions}

For Neumann boundary conditions, we will also consider the regime when the island is located close to the boundary: \mbox{$\rho_I-\rho_0\ll \rho_R$}. In this limit we can also neglect the term $\kappa/\rho_R^{d-1}$. 

Therefore we write the extremum equation in the form
\begin{equation}\label{eq:eqClose}
   \frac{(d-3)}{8G}\frac{\left(\rho_0+(\rho_I-\rho_0)\right)}{r_h^2}=\frac{ c_d}{(\rho_I-\rho_0)^{d-1}}.
\end{equation}
For this equation, there are two possible scenarios to consider: when the boundary is located very close to the island, and when the border is relatively close to the horizon compared to the position of the island.

In the first case we have even more strict constraint \mbox{$\rho_I-\rho_0\ll \rho_0$}, then
\begin{equation}
    \rho_I\approx \rho_0+\left(\frac{c_d}{\frac{(d-3)}{8G}\frac{\rho_0}{r_h^2}}\right)^{1/(d-1)},
\end{equation}
and in the $d=4$ limit and with $G=l_p^{d-2}$ this solution becomes
\begin{equation}\label{eq:eqCloseApprox}
    \rho_I=\rho_0 +\left(\frac{8\,c_d}{\rho_0}\right)^{1/3}l_p^{2/3} r_h^{2/3}.
\end{equation}
In the second case, when the wall is located very close to the horizon, that is, \mbox{$\rho_I-\rho_0\gg \rho_0 $}, we have
\begin{equation}
    \rho_I\approx \rho_0 + \left(\frac{8 c_d}{d-3}\right)^{1/d}l_p^{(d-2)/d} r_h^{2/d},
\end{equation}
and for $d=4$ 
\begin{equation}
    \rho_I=\rho_0+(8 c_d)^{1/4}\sqrt{l_p r_h}\;.
\end{equation}
The found island solutions for the Neumann boundary condition correspond to the local minimum of the generalized entropy.

\section{Quantum extremal surface in fuzzball models}\label{sec:qesFuzz}

The fuzzball paradigm in string theory proposes that a black hole should not be described as a single semiclassical geometry with an event horizon, but rather as an effective coarse-grained description of a large ensemble of horizonless quantum states whose structure extends to the horizon scale \cite{Mathur:2005zp}. Within this framework, a central goal has been to construct smooth geometries in which both the horizon and the singularity are replaced by regular, cap-like structures, allowing for unitary evolution without requiring an interior region hidden behind an event horizon. In the low-energy limit, such configurations are described within supergravity \cite{Bena:2007kg,Bena:2013dka}.

A classic example is the D1--D5--P black hole, which was the subject of the original microscopic entropy calculation by Strominger and Vafa \cite{Strominger:1996sh}. In that setting, the microscopic configurations of branes and strings can be counted in the dual conformal field theory, and their statistical entropy agrees precisely with the Bekenstein--Hawking entropy of the corresponding bulk black hole. This motivates the study of how the island, or quantum extremal surface prescription, behaves when the near-horizon region is replaced by smooth horizonless microstate geometries.

As in the previous sections, the position of the island is determined by extremizing the generalized entropy
\begin{equation}
        S_\text{gen}(\rho_I)=\frac{A(\rho_I)}{4 G}
        -\kappa \frac{A_\perp}{(\rho_R-\rho_I)^{d-2}},
\end{equation}
which leads to the extremality condition
\begin{equation}
    \frac{d S_\text{gen}(\rho_I)}{d\rho_I}
        = \frac{A'(\rho_I)}{4 G} - \kappa (d-2)\frac{A_\perp}{(\rho_R-\rho_I)^{d-1}}=0.
\end{equation}
The first term represents the geometric area contribution, while the second term corresponds to the finite, nonlocal part of the bulk entropy. The existence and stability of an island solution are therefore determined by the interplay between these contributions.

For a Schwarzschild black hole, the near-horizon area exhibits the standard quadratic dependence on proper distance, $A \sim A_h + \rho_I^2$. In the regime $\rho_I \ll \rho_R$, this yields the island position \eqref{eq:Bousso},
\begin{equation}\label{eq:quadrIsl}
    \rho_I\sim \frac{G\; r_h^2}{\rho_R^{d-1}} .
\end{equation}
In this case, $S''_\text{gen}>0$, indicating that the Schwarzschild island corresponds to a stable local minimum of the generalized entropy. For fuzzball geometries, the near-horizon region is instead replaced by a cap or a throat, and the relevant question is whether the corresponding area function can support a similar minimum.

\subsection{D1--D5 and superstrata}

We first consider microstate geometries associated with the D1--D5--P system.  An important class of such solutions is provided by superstrata geometries \cite{Mayerson:2020acj}, which are supersymmetric solutions of six-dimensional minimal supergravity coupled to two tensor multiplets.  The six-dimensional metric can be written as
\begin{equation}\label{eq:superstrata}
    ds_6^2
    =
    -\frac{2}{\sqrt{\mathcal{P}}}(d v+\beta)
    \left(d u+\omega+\frac{\mathcal{F}}{2}(d v+\beta)\right)
    +\sqrt{\mathcal{P}}\; ds_4^2 ,
\end{equation}
where the four-dimensional base is flat $\mathbb{R}^4$.  In spherical bipolar coordinates, the base metric takes the form
\begin{equation}
    ds_4^2
    =
    \Sigma\left(\frac{d r^2}{r^2 + a^2} +d\theta^2\right)
    +(r^2+a^2) \sin^2\theta d\varphi_1^2
    +r^2 \cos^2 \theta d\varphi_2^2,
    \quad
    \Sigma = r^2 + a^2\cos^2\theta,
\end{equation}
and the one-form $\beta$ is given by
\begin{equation}
    \beta\equiv
    \frac{R_y a^2}{\sqrt{2}\Sigma}
    \left(\sin^2\theta d\varphi_1 -\cos^2\theta d\varphi_2\right).
\end{equation}
The warp factor $\mathcal{P}$ is related to the electrostatic potentials by
\begin{equation}
    \mathcal{P}=Z_1 Z_2 - Z_4^2.
\end{equation}

The six-dimensional D1--D5 geometry is obtained from the superstratum ansatz \eqref{eq:superstrata} by setting \cite{Heidmann:2019xrd}
\begin{equation}
    Z_1=\frac{Q_1}{\Sigma},\quad
    Z_2=\frac{Q_5}{\Sigma},\quad
    Z_4=0,\quad
    \mathcal{F}=0,
    \quad
    \omega=\omega_0,
\end{equation}
with
\begin{align}
    \omega_0
    &\equiv
    \frac{a^2 R_y}{\sqrt{2}\Sigma}
    \left(\sin^2\theta d\varphi_1+\cos^2\theta d\varphi_2\right),\\
    \mathcal{P}
    &=\frac{Q_1 Q_5}{\Sigma^2},
    \quad
    Q_1 Q_5=R_y^2 a^2.
\end{align}
With these choices the metric \eqref{eq:superstrata} reduces to the so-called $\text{AdS}_3\times S^3$ supertube geometry,
\begin{align}
    d s_6^2=\sqrt{Q_1 Q_5}\bigg[
    &-\frac{r^2+a^2}{a^2 R_y}dt^2
    +\frac{dr^2}{r^2+a^2}
    +\frac{r^2}{a^2 R_y^2}dy^2 \\
    &+d\theta^2
    +\sin^2\theta \left(d\varphi_1-\frac{1}{R_y} dt\right)^2
    +\cos^2\theta\left(d\varphi_2-\frac{1}{R_y}dy\right)^2
    \bigg].
\end{align}

We now focus on the particular $(1,0,n)$ family of superstrata introduced in \cite{Bena:2016ypk}.  In this case the metric can be written as an $S^3$ fibration over a three-dimensional base manifold $\mathcal{K}$,
\begin{equation}
    ds^2_{(1,0,n)}=ds^2_\mathcal{K}+ds^2_{S^3}.
\end{equation}
The three-dimensional geometry $ds^2_\mathcal{K}$ naturally separates into three regions: a cap region near $r\rightarrow 0$, an intermediate throat region, and an asymptotic region as $r\rightarrow\infty$.

Near the cap, the metric approaches global $\text{AdS}_3$:
\begin{equation}
    \lim_{r\rightarrow 0} ds^2_\mathcal{K}
    =
    \sqrt{Q_1Q_5}
    \bigg[
    \frac{dr^2}{r^2+a^2}
    -
    \frac{a^2 R_y^2}{Q_1^2 Q_5^2}
    \left(a^2+r^2\right)dt^2
    +
    \frac{r^2}{a^2 R_y^2}
    \left(dy +\frac{R_y^2 c}{2 Q_1 Q_5} dt\right)^2
    \bigg].
\end{equation}
On a constant-time slice, after introducing the radial coordinate
\begin{equation}
    r=a \sinh\rho,
\end{equation}
the area of a fixed-$\rho$ surface is
\begin{equation}
    A_\text{cap}(\rho)
    =
    (2\pi \ell \sinh\rho)
    \underset{S^3\text{ Volume}}{(2\pi^2 \ell^3)}
    \approx
    4\pi^3 \ell^4 \rho,
    \quad
    \ell=\sqrt{Q_1 Q_5}.
\end{equation}
Thus, in the cap, the area grows linearly rather than quadratically with the radial proper coordinate.  The generalized entropy and its first two derivatives are therefore
\begin{align}
    S_\text{gen}^\text{cap}(\rho_I)
    &=
    \frac{4\pi^3\ell^4 \rho_I}{4G}
    -
    \frac{\kappa A_\perp}{(\rho_R-\rho_I)^4},\\
    \frac{d S_\text{gen}^\text{cap}(\rho_I)}{d\rho_I}
    &=
    \frac{4\pi^3\ell^4}{4 G}
    -
    \frac{4\kappa A_\perp}{(\rho_R-\rho_I)^5}
    =
    0,\\
    \frac{d^2 S_\text{gen}^\text{cap}(\rho_I)}{d\rho_I^2}
    &<0 .
\end{align}
while the island is located at
\begin{equation}
    \rho_I
    =
    \rho_R
    -
    \left(\frac{4G\kappa A_\perp}{\pi^3 \ell^4}\right)^\frac{1}{5}.
\end{equation} 
However, since the second derivative is negative, this point corresponds to a local maximum of entropy, rather than a minimum. Globally, the entropy value is lower for a trivial island solution, when $\rho_I$ tends towards the cap, i.e., when $\rho_I=0$. Therefore, the quantum extreme surface tends to sit at the cap.

The asymptotic region also approaches $\text{AdS}_3$.  Setting
\begin{equation}
    \rho=r(Q_1 Q_5)^{-1/2}
\end{equation}
as $r\rightarrow\infty$, one finds
\begin{equation}
    \lim_{\rho\rightarrow \infty}ds^2_\mathcal{K}
    =
    \sqrt{Q_1 Q_5}
    \bigg[
    \frac{d\rho^2}{\rho^2}
    +\rho^2 d\tilde{y}^2
    -\rho^2 d\tilde{t}^2
    \bigg].
\end{equation}

The throat region is described by the geometry of an extremal BTZ black hole,
\begin{align}
    ds^2_\mathcal{K}
    &=
    \ell^2
    \bigg[
    \frac{d\rho^2}{\rho^2}
    -\rho^2 dt^2
    +\rho^2 dy^2
    +\rho^2_*(dy+dt)^2
    \bigg],
    \\
    &\rho=r/\sqrt{Q_1 Q_5},
    \quad
    \ell=\sqrt{Q_1 Q_5},
    \quad
    \rho_*=Q_P/(Q_1 Q_5).
\end{align}
This metric is locally $\text{AdS}_3$.  For $\rho\gg \rho_*$ it has the usual $\text{AdS}_3$ asymptotics, while in the near-horizon region $\rho\ll\rho_*$ it can be viewed as a circle of radius $\rho_*$ fibered over $\text{AdS}_2$.  The throat extends over a proper radial length of order $\ell\ln (Q_P/a^2),$ and therefore becomes infinitely deep in the limit $a\rightarrow0$.

On a constant-time slice, the area of a fixed-radius surface behaves as
\begin{equation}
    A_\text{throat}(\rho)\sim\sqrt{\rho^2+\rho^2_*}.
\end{equation}
The proper distance along the throat is
\begin{equation}
    s=\ell \ln\rho,
\end{equation}
so the radial gradient of the area is
\begin{equation}
    \frac{d A_\text{throat}}{ds}
    \sim
    \frac{\rho^2}{\ell\sqrt{\rho^2+\rho^2_*}}.
\end{equation}
Deep in the throat, where $\rho\ll\rho_*$, this gradient is parametrically small.  Consequently, the extremization condition
\begin{equation}
    \frac{1}{4 G}\frac{d A_\text{throat}}{ds}
    -
    \frac{4\kappa A_\perp}{(s_R-s_I)^5}
    =
    0
\end{equation}
can be satisfied only if the radiation region is placed at parametrically large proper distance, $s_R-s_I\rightarrow\infty$.

Thus, within this approximation, the D1--D5 cap and the superstratum throat do not furnish a stable interior island in the same sense as the Schwarzschild near-horizon region.  In the cap, the stationary point is a maximum of the generalized entropy.  In the throat, the area changes too slowly to balance the bulk entropy term except at the edge of a parametrically large separation. Therefore, the potential saddle point is moved to the edge of the assumed region, rather than remaining a robust local minimum.

\subsection{Bubbling solution}

We next consider another class of smooth, horizonless supergravity solutions: five-dimensional bubbling geometries with a four-dimensional Gibbons--Hawking base\cite{Bena:2005va,Bena:2008wt}. The metric takes the form 
\begin{equation}\label{eq:fullBubble}
    ds_5^2
    =
    -\frac{1}{(Z_3\mathcal{P})^{2/3}}(dt+\omega)^2
    +(Z_3\mathcal{P})^{1/3}ds^2_4,
    \quad
    \mathcal{P}\equiv Z_1 Z_2- Z_4^2.
\end{equation}
The base metric $ds_4^2$ is asymptotically flat, with
\begin{equation}
    ds_4^2\rightarrow dr^2+ r^2 d\Omega_3^2.
\end{equation}
The large-radius behavior of the functions $Z_\alpha$, with $\alpha=1,2,3$, determines the conserved charges:
\begin{equation}
    Z_\alpha\rightarrow 1+\frac{Q_\alpha}{r^2},
    \quad
    \alpha=1,2,3.
\end{equation}
Smooth horizonless bubbling solutions are controlled by a length scale $a$.  In the regime with a large hierarchy $a^2\ll Q_\alpha,$ there is an intermediate region
\begin{equation}
    a^2\ll r^2\ll Q_\alpha,
\end{equation}
where
\begin{equation}
    Z_\alpha\simeq \frac{Q_\alpha}{r^2}.
\end{equation}

For vanishing angular momenta, the metric in this region approaches $\text{AdS}_2\times S^3$.  Introducing
\begin{equation}
    \tilde{t}=2t/\sqrt{Q},
    \quad
    \tilde{r}=r^2/Q,
    \quad
    Q\equiv (Q_1 Q_2 Q_3)^{1/3},
\end{equation}
one finds
\begin{equation}
    ds_5^2
    \rightarrow
    \frac{Q}{4}
    \left(
    -\tilde{r}^2 d\tilde{t}^2
    +\frac{d\tilde{r}^2}{\tilde{r}^2}
    \right)
    + Q d\Omega_3^2.
\end{equation}
Therefore, on a constant-time slice, the area of the fixed-$\tilde{r}$ three-sphere is independent of $\tilde r$:
\begin{equation}
    A_\text{throat}=2\pi^2 Q^{3/2},
    \quad
    A'_\text{throat}=0.
\end{equation}
In this exact $\text{AdS}_2\times S^3$ throat region, the area term has zero radial gradient.  The extremization equation therefore has no solution supported purely inside the throat.

To understand what happens near the cap, we use the Gibbons--Hawking form of the base metric,
\begin{equation}
    ds_4^2
    =
    V^{-1} (d\psi+A)^2
    +V ds_3^2,
    \quad
    V\sim\frac{q}{r}.
\end{equation}
On a constant-$t$ slice near the cap, the base geometry is locally
\begin{equation}
    ds_4^2\sim d\rho^2 +\rho^2 d\Omega_3^2,
\end{equation}
where the proper distance behaves as $\rho\sim\sqrt{r}$. Including the warp factor from the full five-dimensional metric \eqref{eq:fullBubble}, the area of a fixed-$\rho$ three-sphere is
\begin{equation}
    A_\text{cap}(\rho)
    \sim
    (Z_3\mathcal{P})^{1/2}
    \underset{S^3\text{ Vol}}{(2\pi^2 \rho^3)}.
\end{equation}
Near the cap, the warp functions scale as
\begin{equation}
    Z_\alpha \sim \frac{1}{r}.
\end{equation}
Since $r\sim \rho^2$, this implies
\begin{equation}
    \begin{aligned}
        \sqrt{Z\mathcal{P}}
        &\sim
        \frac{1}{\rho^3}
        \left(1+C\rho^2 +\mathcal{O}(\rho^4)\right).
    \end{aligned}
\end{equation}
Thus the leading $\rho^3$ behavior of the sphere volume is cancelled by the warp factor, and the cap area takes the form
\begin{equation}
    A_\text{cap}(\rho)\approx A_0+ \text{Const}\cdot\rho^2.
\end{equation}

This is qualitatively different from the D1--D5 cap discussed above.  In the bubbling solution, the area grows quadratically near the cap, just as it does near the Schwarzschild horizon when written in terms of proper distance.  Consequently, the generalized entropy has a saddle point of the same form as \eqref{eq:quadrIsl}, and the resulting saddle is a local minimum.  However, this solution exists only in the region very close to the cap, where the expansion leading to
\begin{equation}
    A_\text{cap}(\rho)\approx A_0+ \text{Const}\cdot\rho^2
\end{equation}
is valid, namely in the regime $r^2\ll a^2$.

\section{Discussion and main results}\label{sec:conclusions}
The main goal of this work was to explore the island prescription and resolving the information paradox in the context of fuzzball-like descriptions of black holes. Since more realistic fuzzball geometries are technically involved, we started with the simplest effective realization of this idea -- replacing the event horizon by a stretched horizon. This setup provides a toy model that captures a key feature of the fuzzball paradigm, namely, the absence of a traditional interior region, while remaining that of where the island prescription could be applied explicitly. 

In the two-dimensional model of Schwarzschild black hole with a stretched horizon, we showed that the inclusion of a reflecting boundary significantly alters both the entanglement entropy and the structure of the island saddles. While the island prescription can, in principle, restore consistency with unitarity, this mechanism is not universally robust in the presence of a boundary. In particular, we observed a “blinking” behavior of the island: for a wall position close enough to the horizon, the island disappears for a finite period of time. During these intervals, the entropy of radiation can again exceed the thermodynamic bound, signaling a reappearance of the information paradox. Thus, even in this simplified setting, the introduction of a boundary does not guarantee the persistence of the island solution.

Considering a two-dimensional approximation of a Schwarzschild black hole with a reflecting boundary near the event horizon, we assume a thermal state at infinity. The calculation of the stress-energy tensor has shown that maintaining a static reflecting wall near the horizon requires an increasingly large external force, diverging in the near-horizon limit. This indicate that such boundary conditions are not dynamically natural within semiclassical gravity.

Extending the discussion to higher dimensions, we incorporated both bulk entanglement contributions and boundary-induced corrections to the generalized entropy. The resulting extremization problem shows that the existence of island solutions is highly sensitive to the boundary conditions and to the position of the stretched horizon. In particular, we found that imposing a boundary can, for a significant region of parameter space, eliminate physically relevant island saddles altogether.

More realistic fuzzball models can be obtained from horizonless microstate geometries, which cap off just above where the horizon of the corresponding black hole would form. However, not all the models we have considered have an island solution. For the existence of a solution to the extremization equations of generalized entropy, the area term included in the entropy must have a quadratic dependence on the position of the island. This is similar to the case of a toy model of a Schwarzschild black hole with a boundary. In the case of a five-dimensional bubbling solution, this occurs only in the region very close to the cap. In contrast, in superstrata geometries, the cap region exhibits only linear growth of the area with proper distance, while the throat region features an extremely small variation of the area term. As a result, any stationary point of the generalized entropy is either a local maximum or pushed to the boundary of the parameter space, preventing the formation of a stable island. These findings suggest that the success of the island prescription in conventional black hole backgrounds relies on specific geometric features that are not automatically reproduced in fuzzball constructions.

This raises an important conceptual point: if fuzzballs provide a fundamentally different description of black holes, then the mechanism by which information is recovered may not always be captured by the standard island framework. Understanding how, or whether, quantum extremal surfaces emerge in such horizonless geometries remains an open question and an important direction for future work.

\acknowledgments
The work  was supported by non-government  Foundation for the Advancement of Theoretical Physics and Mathematics "BASIS" grants \#24-1-3-35-1 (DA).

\newpage
\appendix
\section{Stress tensor calculations}\label{app:stressTensor}
In this section, we will calculate the expectation value of the stress-energy tensor in Herzog and Huang form:
\begin{equation}\label{eq:HerzogHuangApp}
    \langle T^\mu{}_\mu \rangle = \frac{c}{24\pi}\Big(R + 2K\,\delta(x_\perp)\Big).
\end{equation}
With short computation one finds the curvature scalar for the metric above \eqref{eq:metricApp}
\begin{align}
R(r) = - f''(r) \, .
\end{align}
We introduce the proper distance $x_\perp$ orthogonal to the wall,
\begin{equation}\label{eq:perpCoord}
    d x_\perp=\frac{dr}{\sqrt{f(r)}}.
\end{equation}
Thus, at the wall we have
\begin{align}
\delta(x_\perp)\,dx_\perp = \delta(r-r_0)\,dr \qquad \Longrightarrow \qquad \delta(x_\perp) = \sqrt{f_0}\,\delta(r-r_0) \, , \qquad f_0 \equiv f(r_0) \, .
\end{align}
Finally, we compute the extrinsic curvature.  Since the matter region is $r\ge r_0$, the outward normal points toward decreasing $r$.  The outward unit normal is therefore \mbox{$n^\mu = (0,-\sqrt{f})$}.
Using $K=\nabla_\mu n^\mu$ in $(t,r)$ coordinates, one obtains
\begin{align}
K(r) = \partial_r n^r = -\frac{f'(r)}{2\sqrt{f(r)}} \, , \qquad K_0 \equiv K(r_0) = -\frac{f_0'}{2\sqrt{f_0}} \, , \qquad f_0' \equiv f'(r_0) \, .
\end{align}By substituting these components into the Herzog-Huang formula \eqref{eq:HerzogHuangApp}, we obtain the expectation value of the trace of the stress-energy tensor in the $r$ coordinate.
\begin{align}\label{eq:anomalyAppendix}
\langle T^\mu{}_\mu \rangle 
&= \frac{c}{24\pi}\Big(-f''(r) + 2K_0\,\delta(x_\perp)\Big) 
= \frac{c}{24\pi}\Big(-f''(r) - f_0'\,\delta(r-r_0)\Big) \, .
\end{align}
This is the precise meaning in which the anomaly contains a bulk term proportional to $R$ and a boundary-localized term proportional to $K$.

\subsection{Bulk Ward identities and the stationary solution} 

Away from the wall, the stress tensor satisfies the usual bulk Ward identities,
\begin{align}\label{eq:traceBulkAppendix}
\nabla_\mu \langle T^\mu{}_\nu \rangle = 0, 
\quad 
\langle T^\mu{}_\mu \rangle = \frac{c}{24\pi}R = -\frac{c}{24\pi}f''(r), \quad r>r_0 \, .
\end{align}
We now solve these equations explicitly under the mild assumptions appropriate to a static background: stationarity $\partial_t \langle T_{\mu\nu}\rangle=0$ and symmetry $T_{\mu\nu}=T_{\nu\mu}$.  By calculating the covariant derivative and substituting the non-zero Christoffel symbols, we can get the following for the index $\nu=t$:
\begin{align}
0 = \nabla_\mu \langle T^\mu{}_t\rangle  \qquad \Longrightarrow \qquad \partial_r \langle T^r{}_t\rangle = 0 \qquad \Longrightarrow \qquad \langle T^r{}_t \rangle = \Phi \, ,
\end{align}
so the radial energy flux is constant.  For the $r$-component one obtains an equation 
\begin{align}
    0 &= \nabla_\mu \langle T^\mu{}_r\rangle\\
    \partial_r\!\big(f\,\langle T^r{}_r\rangle\big) &= \frac{f'}{2}\,\langle T^\mu{}_\mu\rangle,\quad 
\end{align}
where the trace value can be derived from \eqref{eq:traceBulkAppendix}.

Using $\partial_r\big(\tfrac12 (f')^2\big)=f'f''$, we integrate to find
\begin{align}
\langle T^r{}_r\rangle = \frac{\Omega}{f} - \frac{c}{96\pi}\frac{(f')^2}{f} \, ,
\end{align}
with an integration constant $\Omega$ that encodes the choice of state.  Lowering indices yields the covariant components in the bulk, $r>r_0$,
\begin{align}\label{eq:stressBulk}
\langle T_{tt}\rangle_{\rm bulk} &= \Omega + \frac{c}{24\pi}f f'' - \frac{c}{96\pi}\left(f'\right)^2 \, , \nonumber\\
\langle T_{rr}\rangle_{\rm bulk} &= \frac{1}{f^2}\left(\Omega - \frac{c}{96\pi}\left(f'\right)^2\right) \, , \nonumber\\
\langle T_{tr}\rangle_{\rm bulk} &= \frac{\Phi}{f} \, .
\end{align}
A reflecting wall enforces a vanishing energy flux through the boundary, which sets $T^r{}_t(r_0)=0$ and therefore $\Phi=0$.  In the exterior problem, the horizon is absent from the matter region and regularity at $r=r_h$ does not restrict $\Omega$. Instead $\Omega$ is determined by the asymptotic state at infinity.

\subsection{Distributional conservation and the boundary force (displacement)}\label{app:displacement} 

The presence of a boundary means that normal momentum is not necessarily conserved within the matter region. The lack of normal conservation is measured by the displacement operator. We will demonstrate this explicitly again with a Gaussian normal coordinate $x_\perp\equiv x$\eqref{eq:perpCoord} with $x_\perp=0$ at the wall and $x_\perp>0$ in the exterior region.  In $(t,x)$ coordinates the metric takes the form \mbox{$ds^2=-f(x)\,dt^2+dx^2$}.

The full stress tensor can be written as the sum of a bulk term and a boundary contribution 
\begin{align}
T^{\mu\nu} = \Theta(x)\,T^{\mu\nu}_{\rm bulk} + \delta(x)\,\tau^{\mu\nu}_\text{bdry}\, ,
\end{align}
where $\Theta(x)$ is the Heaviside step function and $\tau^{\mu\nu}_\text{bdry}$ has only one non-zero component $\tau^{tt}_\text{bdry}$, while other components vanish in the minimal realization of the anomaly. From boundary contributions to the stress tensor \eqref{eq:stressBound} we can get
\begin{equation}
    \begin{gathered}
        \langle T_{tt}\rangle_\text{bdry}=\frac{c}{24 \pi}f_0 f_0'\;\delta(r-r_0)=\frac{c}{24\pi}\sqrt{f_0}f_0'\;\delta(x)\\
        \Rightarrow \tau^{tt}_\text{bdry}=\frac{c}{24\pi}\frac{\sqrt{f_0}f_0'}{f_0^2}.
    \end{gathered}
\end{equation}

We can define displacement expectation value using Ward identity \cite{Herzog:2017xha}
\begin{equation}
    \begin{gathered}
        \nabla_\mu T^{\mu x} =\delta(x) D\\
        \nabla_\mu T^{\mu x}=(\nabla_\mu \Theta(x)) T^{\mu x}_\text{bulk}+ \delta(x) \Gamma^x_{tt}\tau^{tt}\;.
    \end{gathered}
\end{equation}
The derivative of $\Theta(x)$ produces a boundary term $\delta(x)\,T^{xx}_{\rm bulk}(0)$, while the second term contributes $\delta(x)\,(\sqrt{f_0} f_0'/2)\,\tau^{tt}$. By replacing variables in the tensor and using equation \eqref{eq:stressBulk}, we get
\begin{equation}
    T^{xx}_{\rm bulk}=f\,T_{rr}^{\rm bulk}=\frac{1}{f}\big(\Omega-\frac{c}{96\pi}(f')^2\big).
\end{equation}
Thus, we find the displacement expectation value 
\begin{align}
D = \frac{\Omega}{f_0} + \frac{c}{96\pi}\frac{f_0'^2}{f_0}\, .
\end{align}
In terms of the proper wall temperature \eqref{eq:OmegaAppendix} and the wall's proper acceleration $a_0$,
\begin{equation}
    \frac{\Omega}{f_0}=\frac{\pi c}{6}T_0^2, \quad a_0 = \frac{f_0'}{2\sqrt{f_0}},
\end{equation}
the displacement expectation value becomes
\begin{align}
    D = \frac{\pi c}{6}T_0^2 + \frac{c}{24\pi}a_0^2 \, .
\end{align}

Additionally, in the orthonormal frame, we can calculate the local pressure and energy density by substituting the previously derived value of $\Omega$
\begin{align}
\rho(r) \equiv \langle T_{\hat t\hat t}\rangle = \frac{\langle T_{tt}\rangle_{\rm bulk}}{f}
= \frac{\pi c}{6}T_{\rm loc}(r)^2 + \frac{c}{24\pi}f''(r) - \frac{c}{96\pi}\frac{(f')^2}{f}\, , \nonumber\\
p(r) \equiv \langle T_{\hat r\hat r}\rangle = f\,\langle T_{rr}\rangle_{\rm bulk}
= \frac{\pi c}{6}T_{\rm loc}(r)^2 - \frac{c}{96\pi}\frac{(f')^2}{f}\, .
\end{align}
The first terms are precisely the local CFT thermal energy density and pressure at the Tolman temperature $T_{\rm loc}(r)$, while the remaining terms encode vacuum polarization dictated by the anomaly.

\subsection{Schwarzschild metric}

In this section, we will calculate specific values for the previously discussed quantities for the two-dimensional Schwarzschild metric with \mbox{$f(r)=1-\frac{r_h}{r}$}. The bulk stress tensor for $r>r_0$ in a reflecting KMS state characterized by $T_0$ becomes
\begin{align}
\langle T_{tt}\rangle_{\rm bulk}
&= \frac{\pi c}{6}\Big(1-\frac{r_h}{r_0}\Big)T_0^2 + \frac{c}{96\pi}\Big(-\frac{8r_h}{r^3}+\frac{7r_h^2}{r^4}\Big)\, , \nonumber\\
\langle T_{rr}\rangle_{\rm bulk}
&= \frac{1}{\big(1-\frac{r_h}{r}\big)^2}
\Big[\frac{\pi c}{6}\Big(1-\frac{r_h}{r_0}\Big)T_0^2 - \frac{c}{96\pi}\frac{r_h^2}{r^4}\Big]\, , \qquad 
\langle T_{tr}\rangle_{\rm bulk}=0\, .
\end{align}
The boundary term required by the anomaly is
\begin{align}
\langle T_{tt}\rangle_{\rm bdry}
= \frac{c}{24\pi}\Big(1-\frac{r_h}{r_0}\Big)\Big(\frac{r_h}{r_0^2}\Big)\delta(r-r_0)\, , \qquad 
\langle T_{rr}\rangle_{\rm bdry}=\langle T_{tr}\rangle_{\rm bdry}=0\, .
\end{align}
The extrinsic curvature and proper acceleration at the wall are
\begin{align}
K_0 = -\frac{r_h}{2r_0^2\sqrt{1-\frac{r_h}{r_0}}}\, , \qquad 
a_0 = \frac{r_h}{2r_0^2\sqrt{1-\frac{r_h}{r_0}}} = -K_0\, ,
\end{align}
and the displacement expectation value takes the explicit form
\begin{align}
D = \frac{\pi c}{6}T_0^2 + \frac{c}{24\pi}a_0^2
= \frac{\pi c}{6}T_0^2 + \frac{c}{96\pi}\frac{r_h^2}{r_0^4\big(1-\frac{r_h}{r_0}\big)}\, .
\end{align}

\newpage
\bibliographystyle{JHEP}
\bibliography{biblio}

\end{document}